\documentclass{aa} 
\usepackage{graphicx}
\usepackage{txfonts}
\usepackage{rotating}
\usepackage{natbib}
\usepackage{longtable}
\def\AMM{NH$_3$}

\def\CO{\mbox{$^{12}$CO}}

\def\WAT{H$_2$O}
\def\METH{CH$_3$OH}
\def\HII{H{\sc ii}}

\def\kms{\mbox{km~s$^{-1}$}}

\def\Vlsr{$V_{\rm LSR}$}

\def\low{{\it Low}}
\def\high{{\it High}}

\begin{document}

\title{Class I and Class II methanol masers in high-mass star forming regions
}
\author{F. Fontani \inst{1,2} \and  R. Cesaroni \inst{3} \and R.S. Furuya \inst{4}
        }
\offprints{F. Fontani, \email{fontani@iram.fr}}
\institute{ESO, Karl Schwarzschild Str. 2, 85748 Garching bei M\"{u}nchen, Germany
        \and
           Institut de Radio-Astronomie Millim\'etrique, 300 rue de la Piscine, Domaine Universitaire, 38406 Saint Martin d'H\`eres, France
           \and
           INAF-Osservatorio Astrofisico di Arcetri, L.go E. Fermi, 5, 50125, Firenze, Italy
           \and
           Subaru Telescope, National Astronomical Observatory of Japan, 650 North A'ohoku Place, Hilo, HI 96720, USA
           } 
\date{Received date; accepted date}

\titlerunning{Methanol masers in high-mass star forming regions}
\authorrunning{Fontani et al.}

\abstract
{Among the tracers of the earliest phases in the massive star formation process, methanol masers have gained
increasing importance. The phenomenological
distinction between Class I and II methanol masers is based on their spatial association 
with objects such as jets, cores, and ultracompact \HII\ regions,
but is also believed to correspond to different pumping mechanisms:
radiation for Class II masers, collisions for Class I masers. }
%
{In this work, we have surveyed a large sample of massive star forming regions in
Class I and II methanol masers. The sample consists of 296 sources, divided into
two groups named \high\ and \low\ according to their [25--12] and [60--12] IRAS colours. Previous studies
indicate that the two groups may contain similar sources in different evolutionary stages,
with \high\ sources being more evolved.
Therefore, the sample can be used to assess the existence of
a sequence for the occurrence of Class I and II methanol masers during
the evolution of a massive star forming region.}
{We observed the 6~GHz (Class II) \METH\ maser with the Effelsberg 100-m
telescope, and the 44~GHz and 95~GHz 
\mbox{(Class I)} \METH\ masers with the Nobeyama 45-m telescope.}
{We have detected: 55 sources in the Class II line (39 \high\ and 16 \low , 12 new detections); 
27 sources in the 44~GHz Class I line (19 \high\ and 8 \low ,
17 new detections); 11 sources in the 95~GHz Class I line (8 \high\ and 3 \low , 
all except one are new detections).
The detection rate of Class II masers decreases with the distance of the source
(as expected), whereas that of Class I masers peaks at $\sim 5$~kpc. This
could be due
to the Class~I maser spots being spread over a region $\la$1~pc, comparable to
the telescope beam diameter at a distance of $\sim 5$~kpc. We also find that
the two Class I lines, at 44~GHz and 95~GHz, have similar spectral shapes,
confirming that they have the same origin.}
{Our statistical analysis shows that the ratio between the detection rates
of Class II and Class I methanol masers is basically the same in \high\ and
\low\ sources. Therefore, both maser types seem to be equally associated with each
evolutionary phase. In contrast,
all maser species (including H$_2$O) have about 3 times higher detection
rates in \high\ than in \low\ sources. This might indicate that the phenomena
that originate all masers become progressively more active with time, during the 
earliest evolutionary phases of a high-mass star forming region.}


\maketitle
%

\section{Introduction}
\label{intro}

Maser lines of water (H$_2$O), hydroxyl (OH), and methanol (CH$_3$OH)
are commonly detected towards regions of
high-mass star formation (e.g. Comoretto et al.~\citeyear{comoretto}; Menten~\citeyear{menten};
Kurtz et al.~\citeyear{kurtz}; Pestalozzi et al.~\citeyear{pestalozzi}). Among these,
methanol masers are the most recently discovered, and their place
in the star formation process is still not well understood.
Menten~(\citeyear{menten}) suggested to classify methanol
masers into two groups: Class I and Class II.
The latter typically coincide in position with hot molecular cores, ultracompact (UC) \HII\ regions,
OH masers and near-IR sources (e.g. Minier et al.~\citeyear{minier};
Ellingsen et al.~\citeyear{ellingsen06}) and are believed to be
radiatively pumped (Sobolev et
al.~\citeyear{sobolev} and references therein; Cragg et al.~\citeyear{cragg05}). 
In contrast, Class~I masers are also found in massive star forming regions,
but usually offset ($\sim 0.1-1$ pc) from other masers, UC \HII\ regions, and
bright infrared sources. They seem to be collisionally pumped (Cragg et al.~\citeyear{cragg})
at the interface between molecular outflows/jets and the quiescent ambient 
material (Plambeck \& Menten~\citeyear{plambeck}). This scenario is
supported by observations that reveal a coincidence
between Class I methanol masers at 44~GHz and
molecular shock tracers (Kurtz et al.~\citeyear{kurtz04}, Voronkov
et al.~\citeyear{voronkov}).  

Even though both species are believed to trace the earliest phases of the massive
star formation process, it is still not clear whether any physical relation exists among them.
Slysh et al.~(\citeyear{slysh})
claim an anticorrelation between the intensity of Class I and II methanol masers in
the same star forming region. On the other hand, Ellingsen~(\citeyear{elli}) searched
for Class~I masers at 95~GHz towards a sample of known 6~GHz Class~II masers and
could not find any (anti)correlation between the two.
It is also poorly understood whether a relationship exists between the occurrence of 
Class I and II masers and the evolution of the corresponding massive star forming 
region. van der Walt~(\citeyear{vanderwalt}) estimated a lifetime of the 6~GHz Class II
methanol maser of a few $10^4$ yrs, thus covering a consistent part of
the early life of a massive star. Ellingsen~(\citeyear{ellingsen06}) 
compared the infrared (GLIMPSE) colours of sources containing Class II methanol 
masers having or not having a Class I methanol 
maser, and found that those associated with Class I masers have redder GLIMPSE colours,
suggesting that the sources hosting Class I masers are less evolved (see also 
Breen et al.~\citeyear{breen}). 
However, until now an evolutionary sequence of methanol masers occurence has not
been well established, partly because Class I 
masers are less studied and models for some Class I lines
show that their excitation is very sensitive to the physical properties of the environment
(Pratap et al.~\citeyear{pratap}). 

To investigate the existence of a sequence for the occurrence of Class I and II 
methanol masers along the evolution of a massive star forming region, one needs to 
study large samples of high-mass young stellar objects (YSOs) 
believed to be in different evolutionary stages. 
A large sample of massive YSO candidates with this property was
identified by Palla et al.~(\citeyear{palla}):
the sources, all with $\delta \geq -30^{o}$, were selected on the basis of 
their large luminosities ($>10^3~L_\odot$) and FIR colours typical of dense
molecular clumps. 
These have been divided into two subsamples based on the IRAS
colours\footnote{We define the IRAS colours as
[$\lambda_{1}-\lambda_{2}$] = $\log_{10}(F_{\lambda_1}/F_{\lambda_2})$}:
sources with $[25-12]>0.57$ and $[60-12]>1.3$ 
were named \high, the others \low. 
The threshold was taken from Wood \& Churchwell~(\citeyear{wec}), who suggested that
UC \HII\ regions have IRAS colours above these limits.
Palla et al.~(\citeyear{palla}) searched for 22~GHz \WAT\ masers towards 
\high\ and \low\ sources, and found a higher detection rate 
in \high\ sources. In the last decade a series of studies 
aimed at investigating the environment associated with such sources
(association with ammonia cores, centimeter and (sub-)millimeter
continuum emission, \CO\ outflows: Molinari et al. 1996; 
Molinari et al. 1998a; Molinari et al. 2000; Brand et al. 2001; Zhang et al. 2001,~2005) 
indicated that \low\ and \high\ sources are
massive stars in a very early evolutionary stage, with the \low\ group
being dominated by the youngest sources. Two prototypical examples of \high\ and 
\low\ objects are respectively IRAS\,20126+4104 (Cesaroni et al. 1997; Cesaroni 
et al. 1999) and IRAS\,23385+6053 (Molinari et al. 1998b; Fontani et al.~2004;
Molinari et al.~\citeyear{mol08a}). 

In this work, we have searched for
6~GHz (Class II), 44~GHz (Class I) and 95~GHz (Class I) 
 \METH\ maser emission in the \high\ and \low\ sources of the
Palla et al.~(\citeyear{palla}) sample, with the aim to find a possible
evolutionary sequence for the occurrence of the different masers. 
The observations were made with the Effelsberg 
100-m telescope (at 6~GHz) and Nobeyama 45-m telescope (at 44 and 95~GHz). 
In Sect.~\ref{obs} we give an overview of the
observations performed and of the data reduction procedure
adopted. The results are presented in Sect.~\ref{res} and discussed
in Sect.~\ref{discu}. A summary of the main findings of this work is given in Sect.~\ref{conc}.

\section{Observations and data reduction}
\label{obs}

\subsection{Effelsberg 100-m telescope observations and data reduction}
\label{eff}

The $5_1 - 6_0$~A$^+$ Class~II methanol maser line at 6.6685192~GHz was observed
with the Effelsberg 100-m antenna towards 296 sources (149 \high\ and 
147 \low; see Table~\ref{tab_det}) in May~2003. These consist of all the 260 sources selected
by Palla et al.~(\citeyear{palla}), plus 36 sources with FIR colours characteristic
of \high\ and \low\ but rejected by Palla et al.~(\citeyear{palla}) because associated
with known \HII\ regions. Data were obtained using two spectral windows 
corresponding to the right- (RCP) and left- (LCP) circular polarization.
Each spectral window had a $\sim 500$ \kms\ bandwidth, with spectral
resolution of $\sim 0.11$ \kms\ (4096 channels). For sources previously
detected in ammonia by Molinari et al.~(\citeyear{mol96}), we centered the 
windows at the Local Standard of Rest (LSR) velocity, \Vlsr, obtained from the
\AMM\ observations. For the other sources the bandwidth was centered at
0~\kms. The half power beam width at the frequency of the \mbox{($5_1 - 6_0$ A$^+$)} 
methanol line is $\sim$120\arcsec. 
Since, due to a technical problem, the two polarisations had remarkably different system temperatures
(45 K for the LCP and 120 K for the RCP), 
in the analysis we used only the LCP spectra.
The conversion factor between main beam brightness temperature and flux density
is 1.45~Jy/K.
The spectra have a typical 3$\sigma$ rms of $\sim$0.15~Jy, comparable to the high
sensitivity survey of 6~GHz methanol masers performed by Caswell et al.~(\citeyear{caswell}).
The pointing was checked towards W3OH and 3C123, and the pointing accuracy was always 
better than 20\arcsec . 
The flux density scale was calibrated by observing NGC7027, whose
flux was assumed equal to 5.9~Jy and 
the flux calibration uncertainty is estimated to be of the order of $\pm 10\%$.

The data were reduced and analysed by means of
the CLASS software, which is part of the GILDAS
package\footnote{The GILDAS software is available at http://www.iram.fr/IRAMFR/GILDAS}
developed at IRAM and Observatoire de Grenoble. 


\subsection{Nobeyama 45-m telescope observations and data reduction}
\label{nobe}

Observations of the Class~I 44.069476~GHz ($7_{0}-6_1$ A$^+$) and 95.169489~GHz (8$_0-7_1$ A$^+$) 
methanol lines towards 88 of the sources observed in the 6~GHz line 
were performed with the Nobeyama Radio Observatory (NRO) 45-m telescope in April 2002.
The 88 sources were randomly selected among the 260 sources of Palla et al. (1991),
depending on the available observing time, and
should have not introduced any bias in the sub-sample.
We used two SIS receivers simultaneously, centered at the rest frequencies
of the lines. The half power beam width was $\sim$38\arcsec\ at 44~GHz and
$\sim$18\arcsec\ at 95~GHz. 
The observations were made in position-switching
mode. The telescope pointing was checked observing nearby
SiO maser sources at 43~GHz; we estimated a pointing
accuracy of 3\arcsec .
The main beam efficiencies were 0.75
and 0.53 at 44 and 95 GHz, respectively. As a backend, we used the AOS-H which
provides velocity resolutions of $\sim 0.14$~\kms\ at 44 GHz
and $\sim 0.06$~\kms\ at 95 GHz. The AOS-H arrays had velocity
coverages of 272~\kms\ at 44~GHz and 126~\kms\ at 95~GHz,
and were centered at the same velocity used for the 6 GHz
observations. The system temperature was
$\sim$300--500~K at 44~GHz, and 700--1300~K at 95~GHz.
All the spectra were calibrated with the
standard chopper wheel method.
Assuming that the telescope beam is Gaussian, the conversion 
between main beam brightness temperature, $T_{\rm MB}$, and flux density,
$F_{\nu}$, is $F_{\nu}({\rm Jy}) = 2.23 \, T_{\rm MB}({\rm K})$.
We reduced the data using the NEWSTAR package
developed by the NRO.
Subsequently, the fully reduced spectra were converted into
CLASS format for further scientific analysis.

Towards some of the sources observed in the
\METH\ lines, we have also obtained spectra of the ammonia (1,1), (2,2), and (3,3) inversion
transitions. The details of these \AMM\ spectra are not presented
in this work; we use these spectra only to derive the systemic LSR velocities,
$V_{\rm sys}$, of the sources whose $V_{\rm sys}$ cannot be found in the literature
(see Tables~\ref{par_classII_high} and~\ref{par_classII_low}, and Sect.~\ref{velocities}).

\section{Results}
\label{res}

In this section we present the detections in the three lines
observed, and the parameters obtained from the fitting procedure
described in Sects.~\ref{eff} and~\ref{nobe}. 
Spectra of all the detected sources are shown in Appendix~A.
The spectra are grouped as follows: sources detected in all the three
lines are shown in Fig.~\ref{spectra1}; sources detected only in the 6~GHz and 44~GHz 
lines are shown in Fig.~\ref{spectra2}; 
sources detected only in the 44~GHz and 95~GHz lines
are shown in Fig.~\ref{spectra3}; sources detected only in the 6~GHz line are shown in
Figs.~\ref{spectra4} and~\ref{spectra5}; sources detected only in the 44~GHz line
are shown in Fig.~\ref{spectra6}. 
The list of detection and non-detections as well as the main parameters of the
lines are tabulated in Appendix~B.

\subsection{New detections}
\label{new}

Our survey has been cross-correlated with catalogs and large surveys of
both Class I and II methanol masers. For the 6.7~GHz line, we used the
catalog of Pestalozzi et al.~(\citeyear{pestalozzi}),
and the recent works of Cyganowski et al.~(\citeyear{cyganowski}) and
Caswell et al.~(\citeyear{caswell}). For the Class I
lines we used: the catalog of Val'tts \& Larionov~(2007);
Chen et al.~(\citeyear{chen}); Pratap et al.~(\citeyear{pratap}); 
Cyganowski et al.~(\citeyear{cyganowski}). We have found 12 new 6.7~GHz
methanol masers, 17 new 44~GHz methanol masers (among which 3 marginal
detections). To our knowledge, all the sources detected in the 95~GHz methanol
masers, except IRAS\,21391+5802 (detected in Val'tts et al.~\citeyear{valtts95}),
are new detections. 

The new detections are marked in italics in Table~\ref{tab_det}.

\subsection{Observed sources and detection summary}
\label{detection}

The observed sources are listed in Table~\ref{tab_det}. The IRAS name, the
type (\high\ = H; \low = L), and the corresponding equatorial 
coordinates are reported in columns 1 to 4.
In column~5 we give the velocities used for the observations (see Sect.~\ref{eff}).
In columns 6 to 8 we give the following information: source detected (Y), undetected (N),
or not observed (--) in the three \METH\ lines (6~Ghz in column~6, 44~GHz in column~7, 95~GHz in 
column~8). For the sake of completeness, in column~9 we give the same information
for the 22~GHz \WAT\ maser emission. For this latter, the information was taken from:
Palla et al.~(\citeyear{palla}), Valdettaro et al.~(\citeyear{valdettaro}), 
Wouterloot et al.~(\citeyear{wouterloot}), Henning et al.~(\citeyear{henning}),
and Medicina archival data.

\begin{figure}
\centerline{\includegraphics[angle=0,width=9cm]{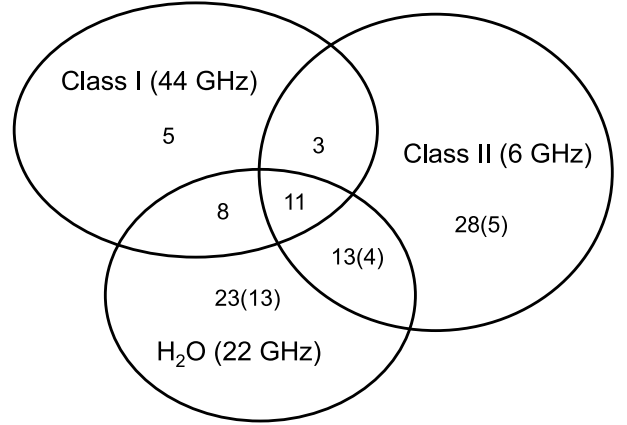}}
\caption{Sketch showing the number of sources detected in the three
maser lines. In parentheses we give the number of sources detected in the
22~GHz \WAT\ maser and/or in the Class II \METH\ maser also observed
in the Class I (44~GHz) \METH\ maser.}
\label{diagram}
\end{figure}

In Fig.~\ref{diagram}, we plot a sketch summarising the number of
sources in which one, two, or all three lines have been detected.  Given that
the sample observed in the Class I \METH\ masers is much smaller than that
observed in the other two lines, for the \WAT\ and Class II \METH\ masers we
also give the number of detected sources that have been {\it observed} also in the
Class I maser (numbers in parentheses in Fig.~\ref{diagram}).

The number of sources observed and detected in the three maser species,
and the corresponding detection rates, are summarised in Table~\ref{rates}. 
We also give the same information for the 22~GHz \WAT\ maser (Col. 11)
and some detection rate ratios: 44GHz / 6GHz (Col.~12), \WAT\ / 6GHz (Col.~13)
and 44GHz / \WAT\ (Col.~14).
For the Class I \METH\ maser line, we consider
the 44~GHz line only because all sources detected at 95~GHz have also
been detected at 44~GHz. 
One can notice that the detection rates of the
Class II masers are comparable to those of \WAT\ masers both
in \high\ and \low\ sources (respectively, 27$\%$ and 28$\%$ in \high\ sources,
11$\%$ and 9$\%$ in \low\ sources), whereas the Class~I masers have a 
higher detection rate than Class~II and \WAT\ masers, in both source types
(48\% in \high\ sources, 17\% in \low\ sources). 

The \high/\low\ detection ratio given in Table~\ref{rates} has the purpose
to highlight possible differences between
the relative occurrence of masers of Class I and II in \high\ and \low\ sources:
The most evident result from Table~\ref{rates} is that the detection rates in all the masers 
(Class I, Class II, \WAT ) are greater in \high\ than in \low\ sources.
On the other hand, their ratios seem to be independent
of the distinction between \high\ and \low\ (Cols.~12, 13 and 14). Therefore,
our analysis does not reveal any significant difference in the {\it relative} occurrence
of the different masers observed in \high\ and \low\ sources. We propose an interpretation
of this result in Sect.~\ref{discu_detection}. 

The 95~GHz Class I line has the lowest total detection rate ($13\%$) and
all the sources detected in this line have been
detected also in the Class I 44~GHz line. We have checked whether the 
different detection rates in the two Class I masers (31$\%$ at
44~GHz versus 13$\%$ at 95~GHz) could be due to worse signal-to-noise
ratio (S/N) at 95~GHz than at 44~GHz.
For this purpose, we have first estimated the noise level that one should
have had in the 44~GHz spectra to achieve the same average S/N
as in the 95~GHz spectra: this is 0.7--0.8 Jy. Then, we have computed the detection
rate that we would have achieved at 44~GHz with this noise: 
this turns out to be 25$\%$, which is still significantly higher than the 13\%
obtained for the 95~GHz line. Therefore, the different detection rates
in the two lines cannot be attributed only to the different noise level,
but must be due to different intensities of the two masers.


\setcounter{table}{0}
\begin{table*}
\begin{center}
\caption{Sources observed (O), detected (D) and corresponding
detection rates (DR) of \METH\ masers, and \WAT\ 
maser at 22~GHz in {\it High} and {\it Low} sources.}
\label{rates}
\begin{tabular}{cccccccccccccc}
\hline \hline
        &  \multicolumn{3}{c}{6~GHz } &  \multicolumn{3}{c}{44~GHz } & \multicolumn{3}{c}{95~GHz } & \WAT\  & 44GHz / 6GHz & \WAT\ / 6GHz & 44GHz / \WAT\ \\
\cline{2-4} \cline{5-7} \cline{8-10}
        &  O  &  D  & DR  &  O  &  D  & DR  & O  &  D  & DR  \\
\hline
{\it High} & 147 & 39  & 27$\pm 4 \%$  & 40 & 19 & $48 \pm 8 \%$  & 40  & 8 & $20\pm 6 \%$ & $28 \pm4\%$ & 1.8$\pm$0.6 & 1.0$\pm$0.3 & 1.8$\pm0.5$  \\
{\it Low}  &  149 & 16 & 11$\pm 3 \%$ &  48 & 8 & $17\pm 5 \%$    & 48   & 3 & $6 \pm 3 \%$ & $9 \pm2 \%$ & 1.5$\pm 0.8$ & 0.8$\pm0.4$ & 1.9$\pm1.0$   \\
\hline
total & 296 & 55 & $19\pm 2\%$ & 88 & 27 & $31\pm 5 \%$ & 88 & 11 & $13 \pm 4\%$  & & & & \\
\hline
{\it High}/{\it Low} & & & 2.9$\pm$1.3 & & & 2.5$\pm1.1$ & & & 3.3$\pm 1.4$ & 3.1$\pm$1.1 & 1.2$\pm$1.0 & 1.2$\pm$1.0 & 0.9$\pm$0.7  \\
\hline
\end{tabular}
\end{center}
\end{table*}


\subsection{Parameters of the 6 GHz (Class II) \METH\ masers}
\label{6ghz_par}

In Table~\ref{par_classII_high} we list the parameters obtained 
from Gaussian fits to the 6~GHz maser lines for \high\ sources.
In column~2 we give the systemic velocity (\Vlsr ) of the sources. 
This has been obtained mainly from \AMM\
observations\footnote{For most sources the velocity
is taken from the \AMM\ survey by Molinari et al.~(\citeyear{mol96}); for 7 sources is 
estimated from \AMM\ spectra obtained with
the Nobeyama 45-m telescope during the same observing
run allocated for the observations of the \METH\ masers (see Sect.~\ref{nobe});
for the remaining sources,
we have taken the systemic
velocities from: Sridharan et al.~(\citeyear{sridharan}); Bronfman et al.~(\citeyear{bronfman},
from observations of the CS(2--1) line); Richards et al.~(\citeyear{richards}) from observations
of HCO$^+$ (1--0); Sunada et al.~(\citeyear{sunada}) and references therein.}.
The best fit parameters are given in columns~3 to 6 as follows: integrated intensity (column~3); 
peak velocity (column~4); line width at half maximum (column~5);
peak intensity (column~6). The same information is given for \low\ sources
in Table~\ref{par_classII_low}.
As can be seen from the spectra in Figs.~\ref{spectra1} -- \ref{spectra5},
in the large majority (85$\%$) of the detected sources the emission consists
in multiple peaks spread over a well defined velocity range, rather
than in a single line.
Interestingly, only 2 out of the 7 sources with only one line in the spectrum
belong to the \high\ group. In fact, 94\% of the detected \high\ sources
have multiple lines, while only 70\% of \low\ sources do.
The line widths are between 0.2 and 0.9 \kms\ in the large majority ($\sim$95\%) 
of the lines, and never exceed 1.5 \kms\ except two cases: 18089--1732, and
one line in the spectrum of 18048--2019. We believe that in these sources (especially 18089--1732, 
for which we derive a line width of 10 \kms) the
large linewidths are due to blending of several narrow lines.
This is suggested by the fact that other surveys of this maser transition in similar IRAS sources
(Szymczak \& Kus~\citeyear{kus}) have revealed line widths smaller than 1~\kms\
in all cases.

In our survey, we find a median flux density peak of $\sim$15~Jy and a median number
of lines per spectrum of 4 (see Tables~\ref{par_classII_high} and
\ref{par_classII_low}). Among the new detections, only two sources have components
with peak flux densities in excess of 6~Jy, IRAS\,05480+2545 and IRAS\,18024--2119, and
10 out of 12 sources have less than 5 components.

Walsh et al.~(\citeyear{walsh97}) observed a sample of 535 IRAS sources south 
of $\delta = 0^{\circ}$ in the 6~GHz \METH\ maser with the Parkes 64-m telescope.
They observed 4 sources that have also been observed in our survey (18159--1648,
18236--1205, 18265--1517 and 18316--0602), all belonging to the \high\ group. 
We find good agreement 
-- both in the flux densities of the lines and in the velocity ranges --
between the spectra taken with the two telescopes in 
all sources, except 18236--1205.
In 18236--1205, the Parkes 64-m observations show more spectral features and
an intensity more than twice the one observed by us with the Effelsberg 100-m telescope.
This is likely due to the variability of the maser line.
In the sources detected in the Parkes survey,
the median value of the flux density peak is $\sim 13$~Jy, and the
median number of lines per spectrum (i.e. per source) is 5 
(Walsh et al.~\citeyear{walsh97}), consistent with our findings. 

\subsection{Parameters of the 44 and 95~GHz (Class I) \METH\ masers}
\label{44ghz_par}

The parameters obtained from gaussian fits to the spectra observed towards
both \high\ and \low\ sources at 44~GHz, shown in
Figs.~\ref{spectra1},~\ref{spectra2},~\ref{spectra3} and~\ref{spectra6}, are
given in Table~\ref{par_classI}.  The structure of the table is the same as
Tables~\ref{par_classII_high} and \ref{par_classII_low} but a letter has been
added to the IRAS name to denote if the source belongs to the \high\ (H) or
\low\ (L) sample.  13 out of the 27
detected objects (i.e.$\sim$50\%) are undetected in the 6~GHz line, among
which the brightest 44~GHz source, 18018--2426.  Contrary to the 6 GHz lines,
at 44~GHz only 8 sources out of the 27 detected show multiple components and
the line widths are generally broader than, or comparable to, 1~\kms;
only 31\% of the sources show linewidths below this limit (see Table~\ref{par_classI}). 
In addition, the lines are not very bright on average:
except 18018--2426, for which the peak flux is
$\sim 350$~Jy, the others have intensities on the order of 1--10~Jy. 
We do not find any systematic differences between the old and new detections 
neither in the peak fluxes nor in the line widths.

The relatively large linewidths suggest either that some of the lines are
thermal, or that they are the superposition of several blended components. We
believe that the second possibility is more likely, on the basis of the
VLA observations of Kurtz et al.~(\citeyear{kurtz04}).
These authors observed some of our sources with better sensitivity and
angular resolution revealing multiple components spread over a velocity range
comparable to the line widths measured by us with the Nobeyama 45-m
telescope.  The sources in common with the survey of Kurtz et
al.~(\citeyear{kurtz04}) are: 18144--1723, 18162--1612, 19092+0841,
21307+5049, 21391+5802 and 23385+6053. All of them have been detected also in
our observations with the sole exception of 21307+5049, which has an intensity peak
of $\sim$1~Jy in the observations of Kurtz et al.~(\citeyear{kurtz04}). Since this value
is larger than the 3$\sigma$ rms in our observations, probably the
intensity of this maser has decreased with time.
%
%
In 18144--1723, 19092+0841 and 23385+6053 we detect only one line, while Kurtz 
et al.~(\citeyear{kurtz04}) detected multiple lines separated by about 2--3~\kms. 
This can explain the line widths observed in our study ($\sim 1.5$, 5.7 and 3.2 \kms, 
respectively). We thus conclude that most of the line widths $>$1~\kms\ are likely
due to multiple components having small ($\leq 0.5$ \kms ) velocity separations. This is 
also supported by the study of Slysh et al.~(\citeyear{slysh}), who detected the 44~GHz line in
148 massive young stellar objects similar to those studied in this work, and
derived linewidths $\la$1~\kms\ in the large majority (87\%, against our 31\%) 
of the sources.

Finally, in Table~\ref{tab_par95GHz} we list the line parameters obtained
from gaussian fits to the spectra at 95~GHz. All sources detected in this line have been
detected also in the 44~GHz transition (see Table~\ref{tab_det}). At 95~GHz,
as well as at 44~GHz, 18018--2426 is the 
brightest source. The line widths are distributed around 1--2 \kms,
and the peak intensities are on the order of 1~Jy, with the above mentioned
exception of 18018--2426
($\sim 50$ Jy), suggesting that some of the detected lines might be 
thermal. In the survey of Kalenskii et
al.~(\citeyear{kalenskii}), the linewidths are similar to those measured in our
sources, but the intensities are comparable to or higher than 40 Jy, suggesting 
maser emission.

\begin{figure}
\centerline{\includegraphics[angle=-90,width=8cm]{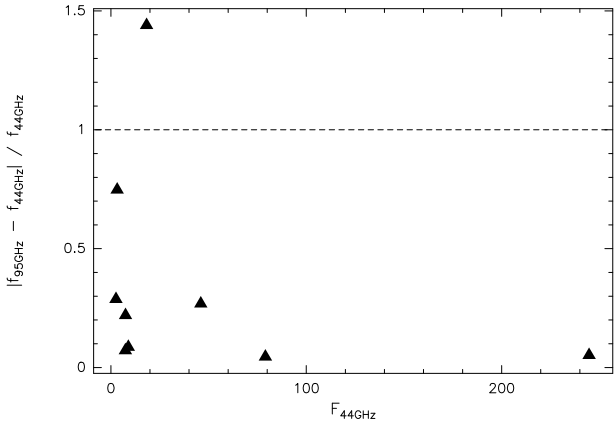}}
\caption{Relative difference between the integrated 
intensity of the lines in the normalised spectra (see text) at 95~GHz and
44~GHz, $|f_{95GHz}-f_{44GHz}|/f_{44GHz}$,  
as a function of the integrated flux $F_{44GHz}$ for 9 out of the 11 sources detected in both lines.
We have not included 18396--0431 because marginally detected in the 44~GHz line, and
21391+5802, for which the 95 and 44~GHz lines have inconsistent
velocity ranges.}
\label{classI_comp}
\end{figure}

At a first glance, the spectra of the 44~GHz and 95~GHz masers shown in
Figs.~\ref{spectra1} and \ref{spectra3} indicate that the shape of these two
lines are similar in a given source, and clearly different from that of the
corresponding 6~GHz maser.  To analyse this effect more quantitatively, we
have normalised each spectrum with respect to the peak intensity and then
calculated the quantity $|f_{95GHz}-f_{44GHz}|/f_{44GHz}$, where $f_\nu$
indicates the integral under the lines of the normalised spectrum.  In
Fig.~\ref{classI_comp} we plot this quantity as a function of the flux
integrated under the 44~GHz lines, $F_{44GHz}$, for 9 out of the 11 sources
detected in both lines.  We have not included 18396--0431,
because the detection at 44~GHz is very doubtful. We have also excluded
21391+5802, for which the 95 and 44~GHz lines have inconsistent
velocity ranges.
Figure~\ref{classI_comp} demonstrates that 7 sources, i.e. $\sim$65\% of the
total, have relative deviations $\leq 0.3$.  Even though the number of
sources is very low, this result indicates that the two Class I masers have
similar shapes in the majority of the sources detected in both lines,
thus confirming the findings of previous studies (e.g.~Val'tts et 
al.~\citeyear{valtts95}).

\section{Discussion}
\label{discu}

An important caveat to be kept in mind when considering our results is that
the angular resolution of our observations (18\arcsec\--120\arcsec)
corresponds to a maximum linear scale of several parsecs. Regions that large
contain a multiplicity of YSOs, possibly in different evolutionary stages, so
that it is impossible to associate our findings for a given maser species with
a single object. Nevertheless, our results have a {\it statistical} significance
in the sense that they represent the {\it mean} properties of the whole star forming
region encompassed by the telescope beam.  The validity of this approach is
demonstrated by the distinction itself between \high\ and \low\ sources, which is
based on the IRAS data, whose resolution is similar to that of our
observations. In this regard, we also stress that such a similarity
guarantees a consistent comparison between our findings and the source
classification based on the IRAS colours.

For these reasons, it is not surprising that we do not find neither a correlation
nor an anti-correlation between the intensity of Class~I and that of Class~II
masers. While this might be expected for different maser types associated
with single YSOs, any such effect is bound to disappear when multiple sources
are observed altogether. On the other hand, one can still use the detection
rates described in Sect.~\ref{detection} to draw statistical conclusions on the relative
evolutionary phase of the different maser types. This will be done in
Sect.~\ref{discu_detection}.

In the following, we will also illustrate and compare properties such as the
typical line width and the line velocity dispersion, for the different maser
types. 


\subsection{Detection rates: \high\ versus \low\ sources}
\label{discu_detection}

The most evident result of this work is that for all maser types the
detection rates are a factor of $\sim 3$ greater in \high\ sources than in
\low\ sources, which indicates that all types of maser emission in a
high-mass star forming region become more easily excited as evolution
proceeds.  This conclusion is confirmed by the fact that, as noted in
Sect.~\ref{detection}, the {\it ratio} between the detection rates of two
different maser types does not change significantly from \high\ to
\low\ sources. In other words, all masers
appear to evolve in a similar way from the \low\ to the \high\ phase, because the
proportion of detected masers remains the same in Low and High sources. 

A possible interpretation of these results is that all phenomena triggering
population inversion and hence maser emission (outflow/accretion shocks,
hot dust emission, etc.) become more and more prominent during the
evolution. Therefore, as time goes on, in a given star forming region
the number of bipolar outflows, accretion disks, and luminous infrared
sources increases, thus leading to a corresponding increase of the number
of maser sources.

Our findings appear to contradict the suggestion by
Ellingsen~(\citeyear{ellingsen06}) that Class I masers are associated
with younger sources than those associated with \METH\ Class II and
\WAT\ masers. However, Ellingsen's claim is admittedly based on a low
number of sources and, as stated by the author himself, is rather speculative.
A larger number of data with better angular resolution for both methanol maser
types will be needed before drawing any conclusion on this issue.

\subsection{Detection in Class I and II masers and source distance}
\label{dist}

\begin{figure}
\centerline{\includegraphics[angle=-90,width=9cm]{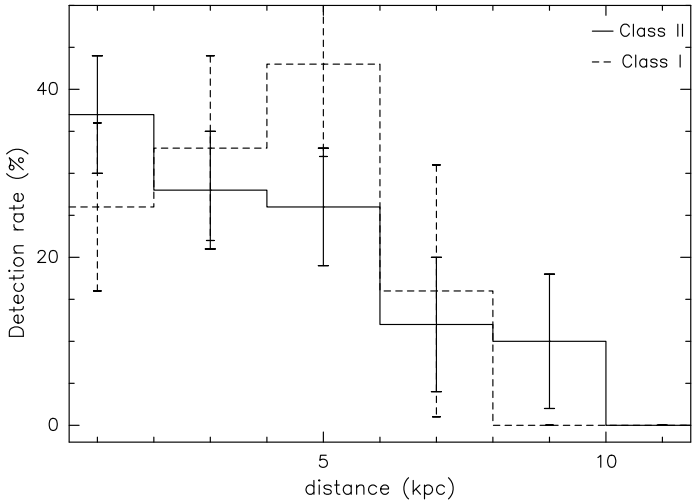}}
\caption{Distributions of the detection rate for the 
Class~II 6~GHz maser (solid line) and Class~I 44~GHz maser 
(dashed line) as a function of the source distance.}
\label{histo_dist}
\end{figure}

In Fig.~\ref{histo_dist} we plot the detection rates
for the \METH\ masers at 6~GHz (Class II) and 44~GHz~(Class I) as a function
of the source distance. The kinematic distance was taken from the following papers:
Molinari et al.~(\citeyear{mol96},~\citeyear{mol08a}); Sridharan et
al.~(\citeyear{sridharan}); Zhang et al.~(\citeyear{zhang05}); Sunada et
al.~(\citeyear{sunada}).
The plot shows that most of the sources detected at 6~GHz
have distances between 1 and 5~kpc and the
detection rate rapidly decreases with increasing distance. 
This is the obvious consequence of the fact that the closer
is the source, the higher is the measured flux. However, while the
Class~II maser detection rate increases all the way to the smallest distances,
that of the Class~I masers attains a maximum at $\sim$5~kpc (45\% of the
sources observed between 4 and 6~kpc are detected) and then
decreases for smaller distances, despite the fact that the 
distance distribution of the whole sample peaks at distances lower than 
$\sim 5$ kpc (see Fig.~3 in Molinari et al.~\citeyear{mol96}).
Our interpretation is that in each source the Class~I maser spots
are spread over a significantly larger region than that of the Class~II maser
spots.
Consequently, in the closest objects part (or all) of the Class~I spots
happen to fall
outside the instrumental HPBW. This idea is supported by the fact that --
as argued in Sect.~\ref{intro} -- Class~I masers are located 
typically 0.1--1~pc away from the centre of star formation activity, 
whereas Class~II masers are found closer to it.
At 44~GHz the HPBW
of the Nobeyama 45-m telescope is $\sim$38\arcsec, which
translates into a linear size of $\sim$0.9~pc at a distance of 5~kpc,
consistent with the maximum spread observed in Class~I maser spots.


\subsection{Velocity ranges in Class I and Class II masers}
\label{velocities}

\begin{figure}
\centerline{\includegraphics[angle=0,width=9cm]{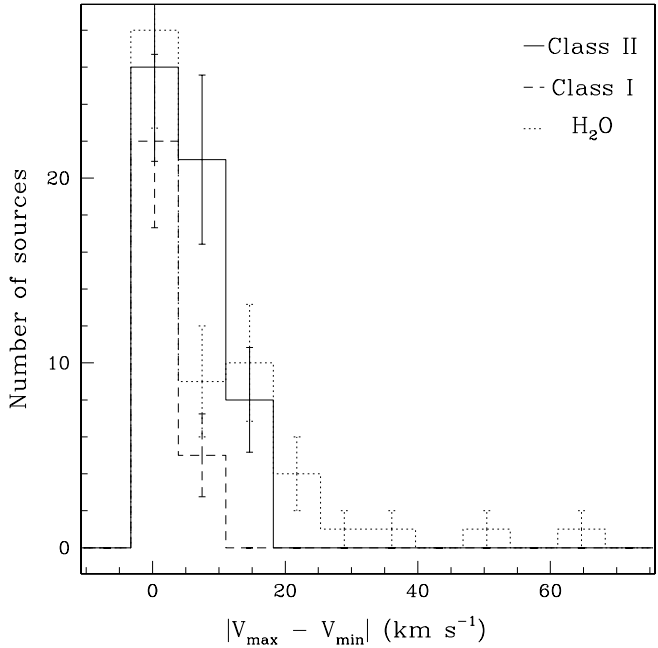}}
\caption{Distribution of the difference between minimum and maximum
line velocities for the Class~II 6~GHz (solid line), Class~I 44~GHz (dashed line) 
methanol masers, and of the 22~GHz \WAT\ maser (dotted line). For spectra with
a single line, we plot the measured full width at half maximum. }
\label{histo_velrange}
\end{figure}

Figure~\ref{histo_velrange} shows the distribution of the velocity range,
derived as the difference between the minimum and maximum velocity peak
of the different lines in each spectrum, $|V_{\rm max}-V_{\rm min}|$, for the
Class~II 6~GHz and Class~I 44~GHz \METH\ masers, and for the sources
detected in the 22~GHz \WAT\ masers by Palla et al.~(\citeyear{palla}; see their
Table~2a) and Valdettaro et al.~(\citeyear{valdettaro}; Table 1). Out of the 55 sources
detected in the \WAT\ maser line, 44 belong to the Palla et al.~(\citeyear{palla})
sample.
For spectra with a single line, we have assumed as velocity range the measured full width
at half maximum (Col.~5 of Tables~\ref{par_classII_high}, \ref{par_classII_low}
and \ref{par_classI}), to be consistent with Palla et al~(\citeyear{palla}).
The majority of the sources show maser emission of Class I \METH\ spread over velocity 
ranges smaller than 6~\kms\ (93\%), while 60\% and $\sim$64\% of the \WAT\
and Class II maser emission, respectively, have a velocity range smaller
than this value, suggesting once more that Class II masers originate from an
environment different from that required to excite Class I masers.
The distributions of the velocity ranges for Class II and \WAT\ masers look
more similar each other. However, that of the Class II masers extends up to velocity
ranges of $\sim 12$~\kms, whereas the distributions of the Class I and \WAT\ 
masers are clearly more peaked at small values: Fig.~\ref{histo_velrange} 
shows that 21 out of 55 sources (38\%) detected in the Class II maser have
velocity ranges between $\sim$6 and $\sim$12~\kms, wehreas only 16\% and
19\% of the sources detected in the \WAT\ and Class I lines have velocity
ranges in this interval. Finally, the largest
difference between the maximum and minimum velocities is found in \WAT\
masers. In fact, 6 sources show \WAT\ emission spread over a velocity range
larger than $\sim 20$~\kms, whereas in all
\METH\ masers the spread is below this limit. In summary, Fig.~\ref{histo_velrange}
indicates that there are no
clear correlations among the velocity spreads of any of the three maser types.

\begin{figure}
\centerline{\includegraphics[angle=0,width=9.5cm]{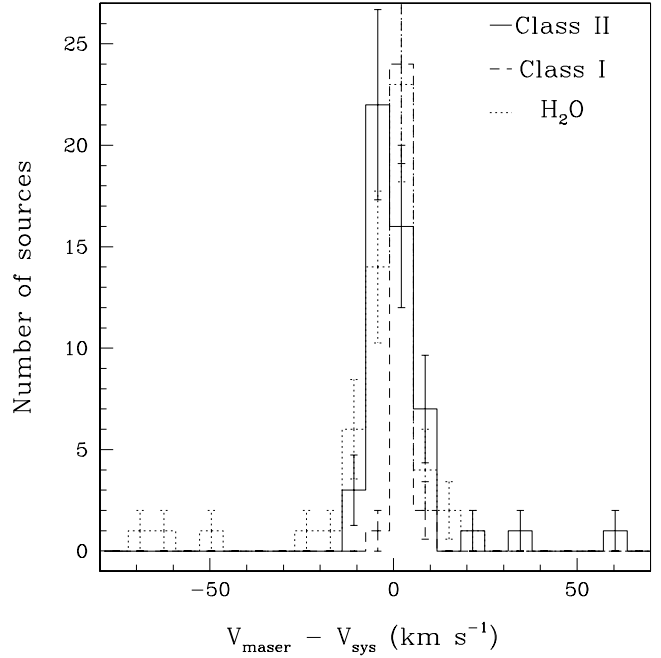}}
\caption{Distribution of the difference between maser
velocity and systemic velocity of the Class II 6~GHz (solid line), Class I 44~GHz (dashed line) 
methanol masers, and of the 22~GHz \WAT\ maser (dotted line).
For sources with multiple lines, we have taken the velocity of the strongest one.}
\label{vsys_vspot}
\end{figure}

In Fig.~\ref{vsys_vspot}, we plot the distribution of the difference
between the velocity of the maser, $V_{\rm maser}$, and the systemic one,
$V_{\rm sys}$ once again for the three maser types. 
Obviously, we did not include in Fig.~\ref{vsys_vspot} sources without available
measurements for $V_{\rm sys}$. These are: 06291+0421, 17582-2234
and 19045+0813, among the sources detected in the \WAT\ maser, and 17571-2328 and 
19186+1440 (see Tables~\ref{par_classII_high} and \ref{par_classII_low}) among
those
detected in the \METH\ masers. $V_{\rm sys}$ for the \WAT\ maser
detections was taken from ammonia measurements (Molinari et al.~\citeyear{mol96},
Valdettaro et al.~\citeyear{valdettaro}).
For spectra with multiple lines we have adopted as $V_{\rm maser}$ 
that of the peak. The 6~GHz 
masers show a distribution centred around zero, but the majority
of the sources ($\sim$67\%) have a separation of $\sim 5$ \kms\ or more
between the systemic and the peak velocity.
On the other hand, for the 44~GHz lines, the distribution is much more peaked around zero
and only 6 sources (22$\%$) have separation larger than $\sim 5$ \kms.
The distribution of the \WAT\ maser resembles more closely  that of the Class II \METH\ line, and it 
shows the largest absolute difference 
between $V_{\rm maser}$ and $V_{\rm sys}$. However, the sources with the largest
velocity separation have $V_{\rm maser}$ blue-shifted with respect to $V_{\rm sys}$, 
while for Class II lines the largest difference is observed in redshifted lines.

Altogether, the results shown in Figs.~\ref{histo_velrange} and~\ref{vsys_vspot}
suggest that the 44~GHz maser emission is originated in gas with smaller
velocity spread and closer to the systemic velocity, than the gas in which
the other masers are produced.  This finding appears in contrast with the
belief that Class I masers could be associated with
jets/outflow (Plambeck \& Menten~\citeyear{plambeck}; Kurtz et
al.~\citeyear{kurtz04}), unless the jet axis lies very close to the plane of the
sky.  Before drawing any conclusion, though, one must take into account the
effect of the different noise at 44 and 6~GHz, which can play a significant
role in shaping the distribution in Fig.~\ref{histo_velrange}. To take this
into account, we proceed as done in Sect.~\ref{detection} for the comparison
between the 44 and 95~GHz spectra.  The 3$\sigma$ rms of the
Effelsberg 100-m spectra is $\sim$0.03--0.04~Jy, whereas the one needed to
achieve the same average signal to noise ratio as at 44~GHz is $\sim$2.2~Jy.
By inspecting Tables~\ref{par_classII_high} and \ref{par_classII_low} one can
see that several lines are below this limit, especially those at the edge of
the velocity interval, i.e.  those used to calculate $|V_{\rm
max}-V_{\rm min}|$. Therefore, the different rms noise at the two frequencies
plays an important role in the estimate of $|V_{\rm max}-V_{\rm min}|$.  To
quantify this effect, we have recalculated the histogram in
Fig.~\ref{histo_velrange} for the 6~GHz line, excluding the lines fainter
than $\sim 2.2$~Jy, and find that the histogram is still slightly wider than
that at 44~GHz. In conclusion, we believe that the different distributions
for the three maser types correspond to different formation environments,
with Class~I masers possibly forming in shocks slower than those generating
e.g. \WAT\ masers and/or propagating
perpendicular to the line of sight.

\section{Summary and Conclusions}
\label{conc}

We have searched for three methanol maser lines (6~GHz Class II, 44~GHz Class I, and
95~GHz Class I) towards a large sample of massive YSOs
divided into two groups (\high\ and \low ) on the basis of their IRAS
colours (Palla et al.~\citeyear{palla}). Previous studies indicate that
the two groups are in different evolutionary stages, with \high\ sources being the 
more evolved. This work aims to use this sample to test a possible evolutionary
sequence in the appearence of Class I and II methanol masers.
The main findings of our study are the following:
\begin{itemize}
\item We have detected: 55 sources in the Class II line (39 \high\ and 16 \low , 12 new detections); 
27 sources in the 44~GHz Class I line (19 \high\ and 8 \low ,
17 new detections); 11 sources in the 95~GHz Class I line (8 \high\ and 3 \low , all new detections).
\item The detection rates of all the masers observed are greater (by a factor
$\sim$3) in \high\ than in \low\ sources: the \high/\low\ detection rate
ratios are 2.9$\pm1.3$ for the Class I 44~GHz line, $3.3\pm1.6$ for the 95~GHz
Class I line, and $2.5\pm 1.1$ for the Class II
line. All these values are similar to that found for \WAT\ masers, i.e. $3.1\pm 1.1$.
Going from \low\ to \high\ sources,
we do not find any statistically significant 
difference in the {\it relative} occurrence of Class I masers with respect to Class II masers.
An analogous result holds for the ratio between the detection rates of CH$_3$OH and H$_2$O masers.
A possible interpretation is that all maser species
analysed in this work evolve similarly during the evolutionary phases
corresponding to \high\ and \low\ sources.
\item The detection rate of the Class II masers decreases with the distance
of the source, as expected, whereas that of the Class I masers peaks at $\sim$5~kpc.
We interpret this result with Class I maser spots being typically
spread over a larger ($\la$1~pc) region than Class II maser spots.
\item The spectra in the two Class I masers appear to have similar shapes
in 7 out of the 11 sources detected in both lines, confirming a common
physical origin. Their different detection rates cannot be explained only
with the different noise levels at 44 and 95~GHz and we thus conclude that
the 95~GHz line is intrinsically fainter.

\end{itemize}

{\it Acknowledgments.} 
We thank the anonymous referee for the constructive criticisms that helped
us to shorten, focus, and substantially improve the paper.
We are grateful to the Effelsberg and
Nobeyama staff for their help during the observations.
For part of this work, FF acknowledges support by Swiss
National Science Foundation grant (PP002 -- 110504). 
The research leading to these results has received funding from
the European Community's Seventh Framework Programme (FP7/2007--2013)
under grant agreement No. 229517.
This work is partially supported by a Grant-in-Aid from the Ministry
of Education, Culture, Sports, Science and Technology of Japan
(No. 20740113).


{}

\clearpage

\renewcommand{\thefigure}{A-\arabic{figure}}
\setcounter{figure}{0}
\section*{Appendix A: Spectra of detected sources}
\label{appa}

\clearpage 

\begin{figure*}
\centerline{
                     \includegraphics[angle=0,width=16cm]{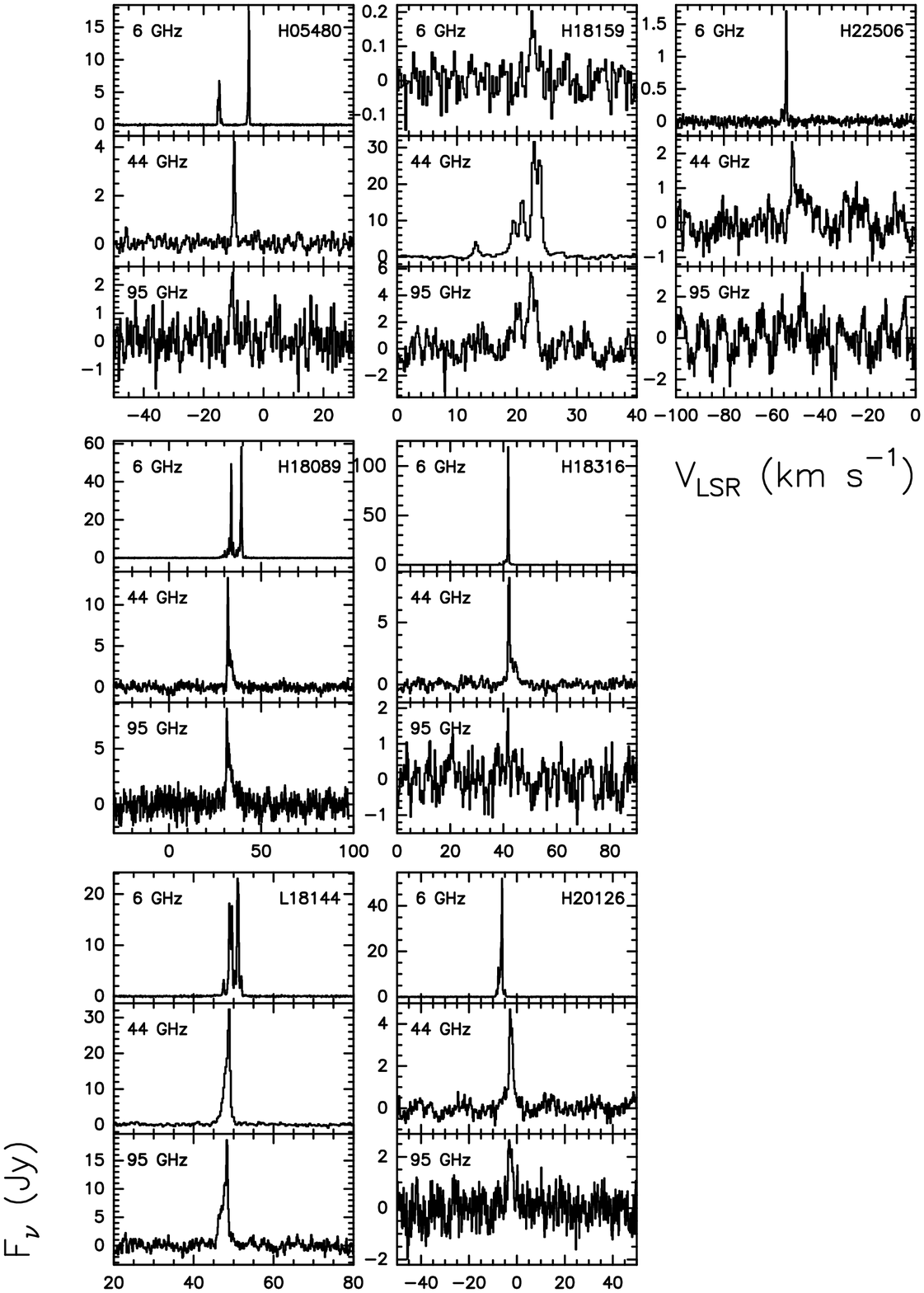}}
\caption[]{Spectra of the three \METH\ masers (from top to 
bottom: 6~GHz, 44~GHz and 95~GHz) observed towards
the 9 sources for which all masers have been detected.}
\label{spectra1}
\end{figure*}

\begin{figure*}[t!]
\centerline{
                     \includegraphics[angle=0,width=16cm]{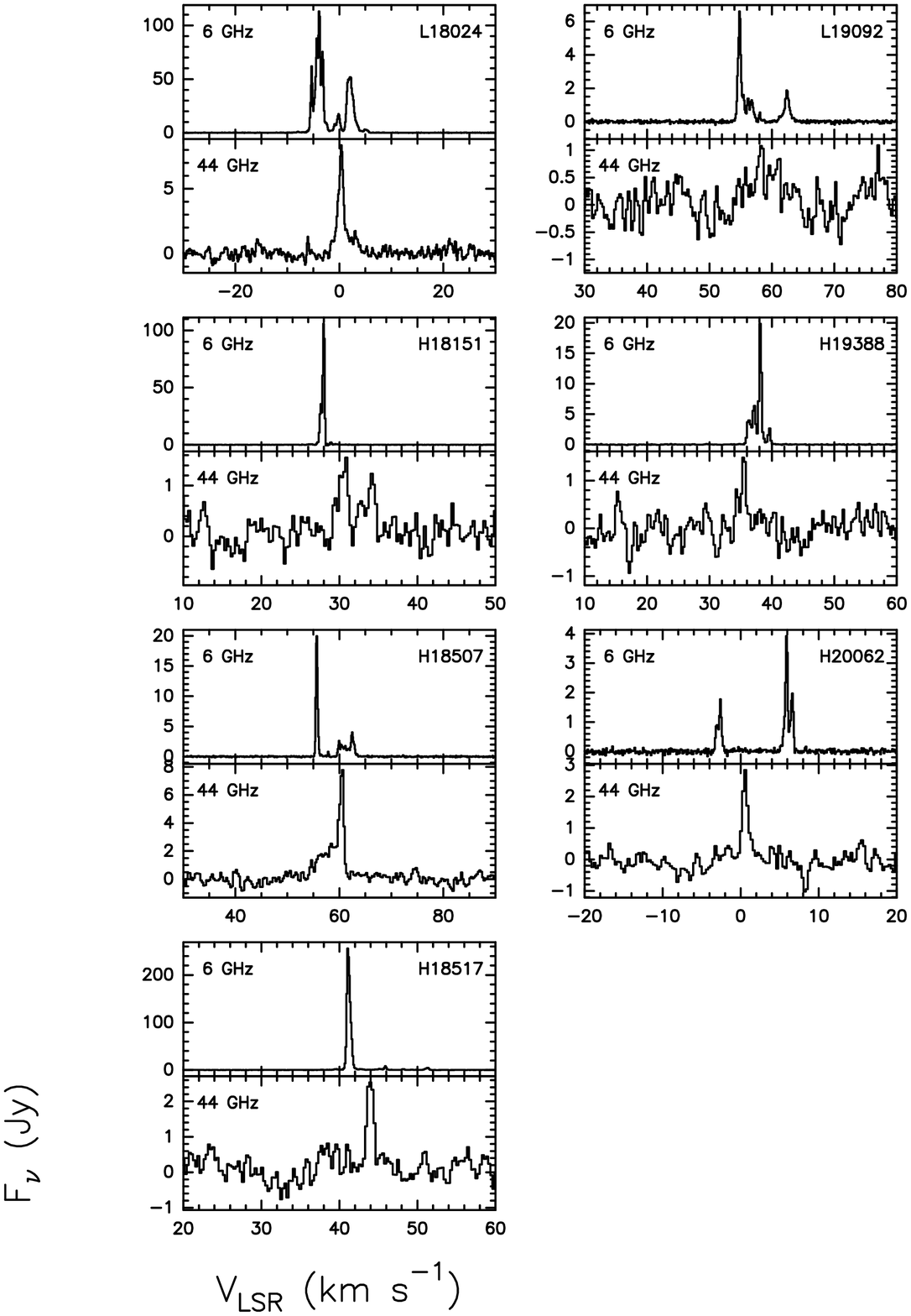}}
\caption[]{Same as Fig.~\ref{spectra1} for sources detected in the
6~GHz and 44~GHz masers.}
\label{spectra2}
\end{figure*}

\begin{figure*}[t!]
\centerline{
                     \includegraphics[angle=0,width=16cm]{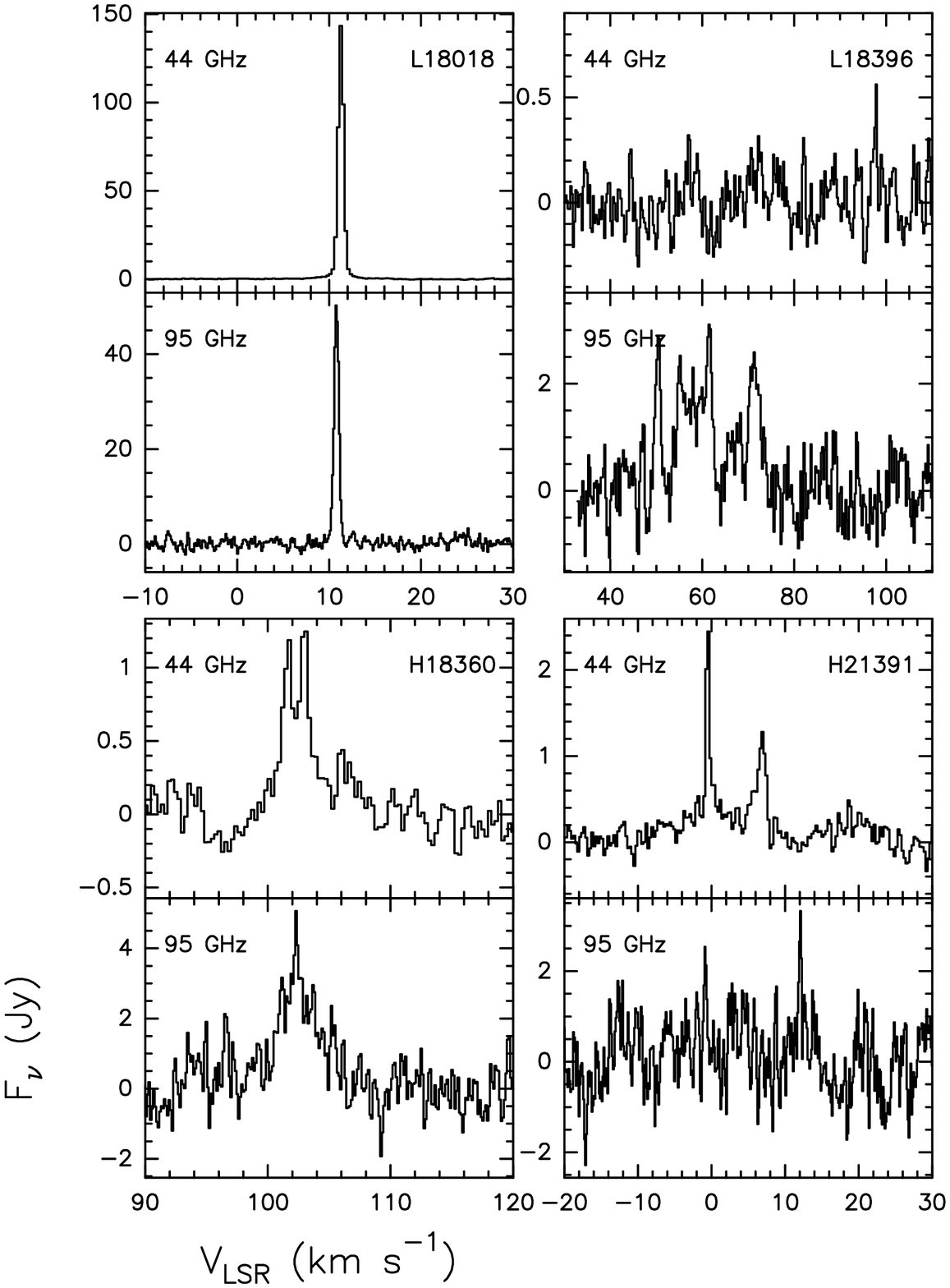}}
\caption[]{Same as Fig.~\ref{spectra1} for sources detected in the
44~GHz and 95~GHz masers.}
\label{spectra3}
\end{figure*}

\begin{figure*}[t!]
\centerline{
                     \includegraphics[angle=0,width=16cm]{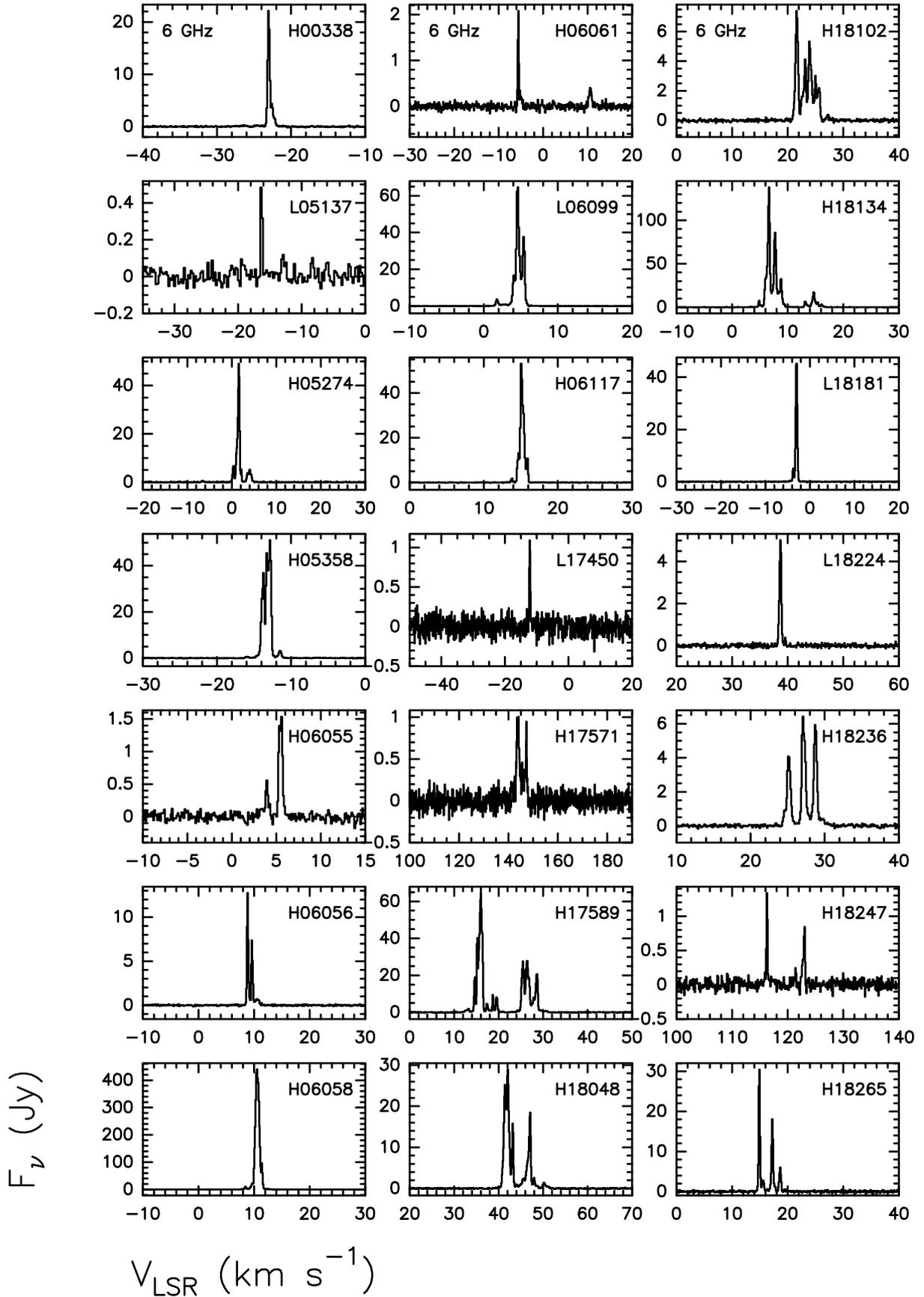}}
\caption[]{Same as Fig.~\ref{spectra1} for sources detected in the
6~GHz maser only.}
\label{spectra4}
\end{figure*}

\begin{figure*}[t!]
\centerline{
                     \includegraphics[angle=0,width=16cm]{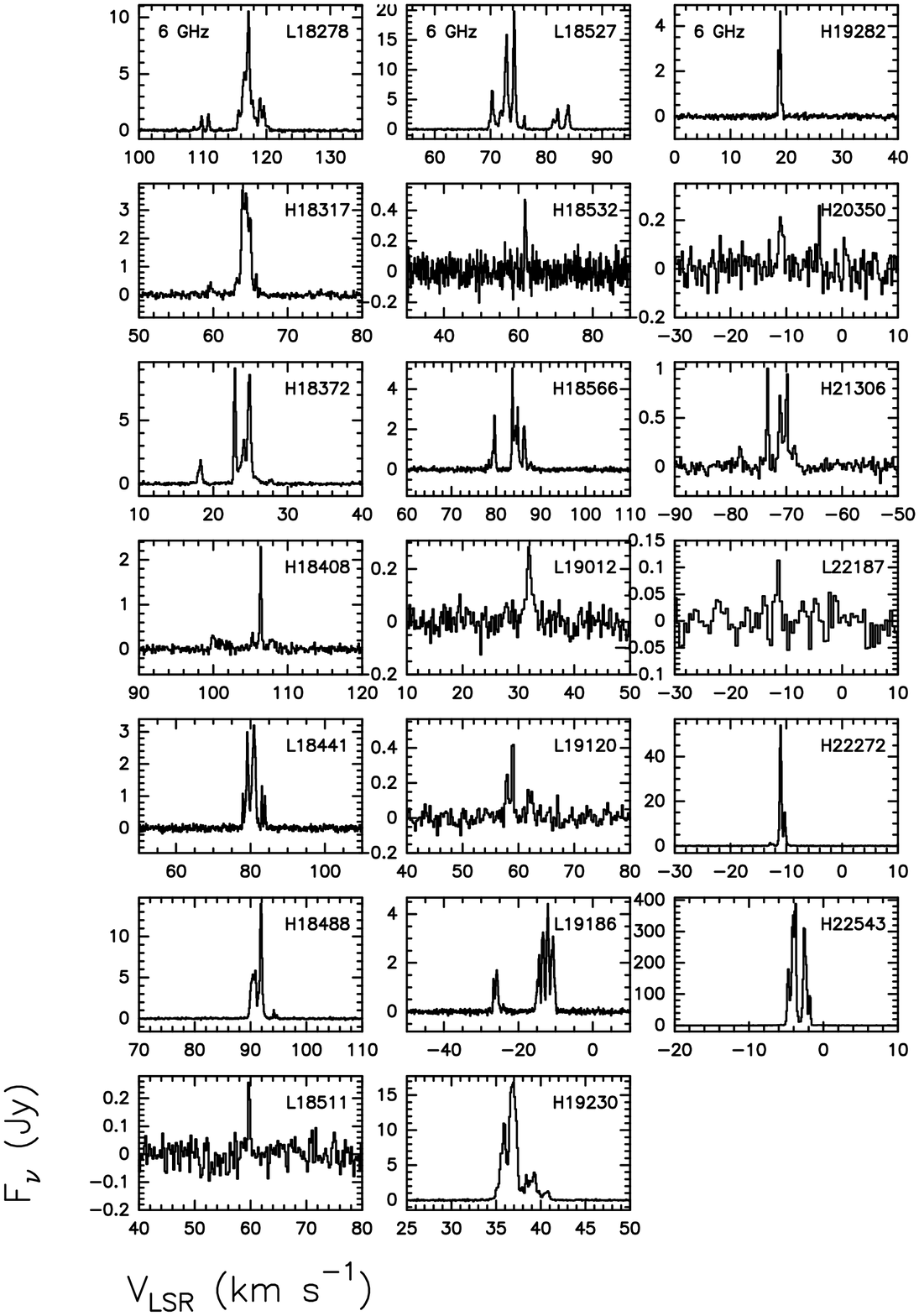}}
\caption[]{Fig.~\ref{spectra4} continued.}
\label{spectra5}
\end{figure*}

\begin{figure*}[t!]
\centerline{
                     \includegraphics[angle=0,width=16cm]{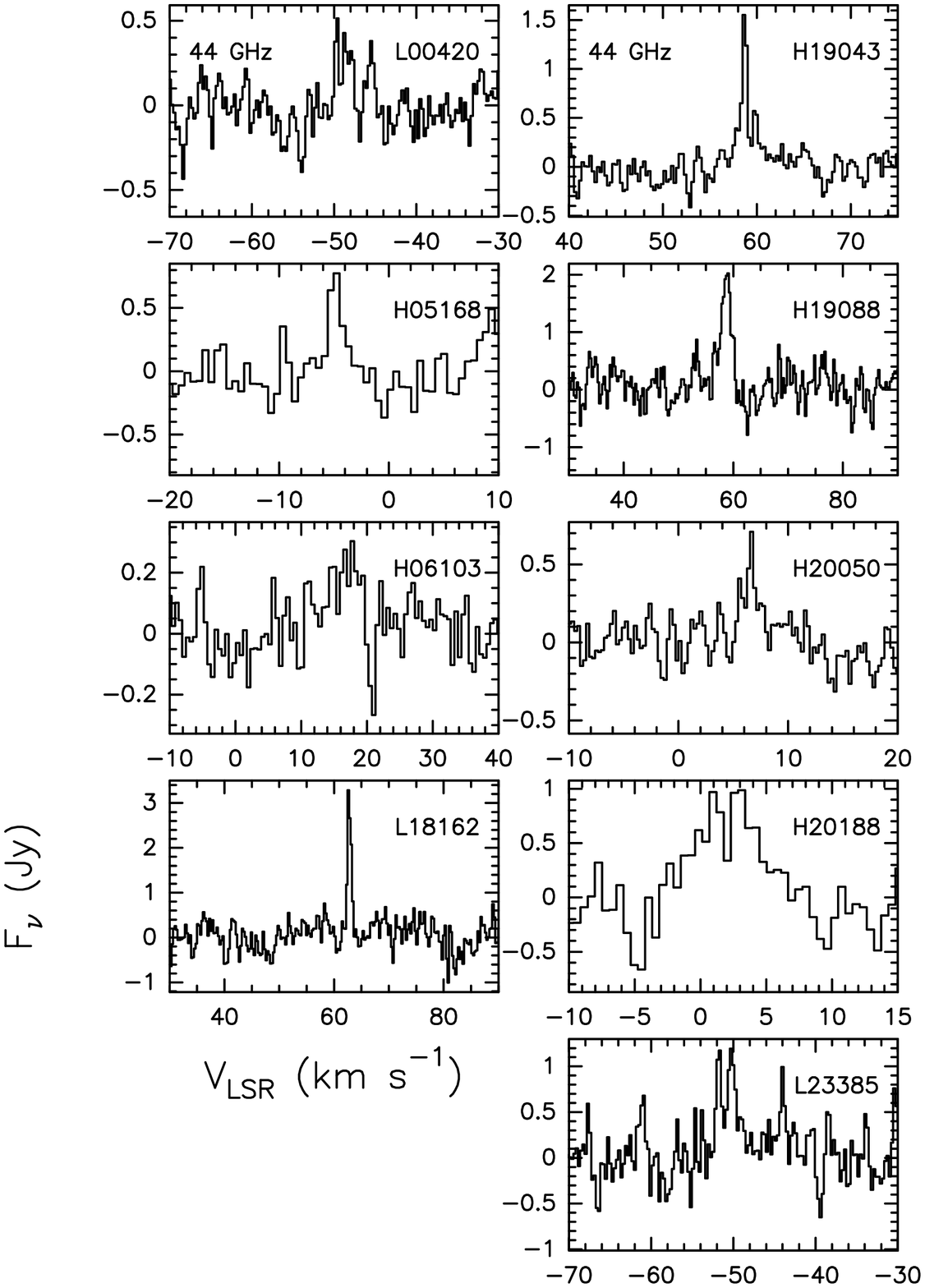}}
\caption[]{Same as Fig.~\ref{spectra1} for sources detected in the 44~GHz line only.}
\label{spectra6}
\end{figure*}


\clearpage

\renewcommand{\thetable}{B-\arabic{table}}
\setcounter{table}{0}
\section*{Appendix B: Tables}
\label{appb}

\clearpage 

\addtocounter{table}{1}

\longtab{1}{
\begin{longtable}{ccccccccc}
\caption{\label{tab_det} Source parameters used for the observations and detection summary. The new detections are in italics.} \\
\hline \hline
IRAS name & type & R.A. (J2000) & Dec. (J2000) &    $V_{\rm LSR}$       & \multicolumn{3}{c}{\METH } & \WAT\ \tablefootmark{p} \\
              &    &                     &                         & \kms\  & 6 GHz & 44 GHz  & 95 GHz  & 22 GHz \\
\hline
\endfirsthead
\caption{continued.} \\
\hline\hline
IRAS name & type & R.A. (J2000) & Dec. (J2000) &    $V_{\rm LSR}$       & \multicolumn{3}{c}{\METH } & \WAT\ \tablefootmark{p} \\
              &    &                     &                         & \kms\  & 6 GHz & 44 GHz  & 95 GHz  & 22 GHz \\
\hline
\endhead
\hline
\endfoot
00070+6503  & L & 00:09:43.7 & +65:20:09 &  0.0  &   N  & -- & -- & N \\
00117+6412  & H & 00:14:27.7 & +64:28:46 &   -36.3  &   N  & -- & -- & N \\ 
00127+6058\tablefootmark{a}   & L & 00:15:29.1 & +61:14:41  &  0.0&   N  & -- & -- & N\tablefootmark{w}  \\
00338+6312\tablefootmark{a}   & H & 00:36:47.5 & +63:29:02  &  0.0  &   Y  & -- & -- &  N\tablefootmark{v}  \\
00420+5530  & L & 00:44:57.6 & +55:47:18   &   -51.2  &   N  & {\it Y} & N &  Y \\
00484+6531  & L & 00:51:33.0 & +65:47:58  &  0.0  &   N  & N & N & N \\
00554+6524  & L & 00:58:39.0 & +65:40:42  &  0.0  &   N  & N & N & N \\
01056+6251\tablefootmark{a}   & L & 01:08:48.5 & +63:07:19  &  0.0  &   N  & -- & -- & N\tablefootmark{w}   \\ 
01420+6401  & L & 01:45:39.6 & +64:16:02  &  0.0  &   N  & -- & -- &  N \\
02383+6241  & L & 02:42:20.1 & +62:53:51  &  0.0  &   N  & N & N  & N \\
02434+6018  & L & 02:47:15.8 & +60:30:44  &  0.0  &   N  & N & N &  N \\
02593+6016\tablefootmark{a}   & H & 03:03:17.9 & +60:27:52  &  0.0  &   N  & -- & -- & N\tablefootmark{v} \\   
03064+5638  & L & 03:10:15.2 & +56:50:18  &  0.0  &   N  & N & N &  N \\
03211+5446  & L & 03:24:59.1 & +54:57:25 &  0.0  &   N  & -- & -- &  N \\
03353+5550  & L & 03:39:15.6 & +55:59:48  &  0.0  &   N  & N & N & N \\
03448+5545  & L & 03:48:47.9 & +55:54:45  &  0.0  &   N  & N & N & N \\
04000+5052  & L & 04:03:49.3 & +51:00:48  &  0.0  &   N  & N & N & N \\
04034+5116  & H & 04:07:11.9 & +51:24:45  &  0.0  &   N  & -- & -- &  N \\
04156+5251\tablefootmark{a}   & L & 04:19:32.7 & +52:58:36  &  0.0  &   N  & -- & -- & N\tablefootmark{w}  \\   
04269+3510\tablefootmark{a}   & L & 04:30:14.4 & +35:16:30  &  0.0  &   N  & -- & -- &  N\tablefootmark{v} \\ 
04579+4703  & H & 05:01:39.7 & +47:07:23 &   -16.5  &   N  & -- & -- & Y \\
05100+3723\tablefootmark{a}   & H & 05:13:25.4 & +37:27:04  &   0.0  &   N  & -- & -- & N\tablefootmark{w}   \\
05137+3919  & L & 05:17:13.3 & +39:22:14 &   -25.4  &   {\it Y}  & N & N &  Y \\
05168+3634  & H & 05:20:16.1 & +36:37:21 &   -15.1  &   N  & {\it Y\tablefootmark{m}} & N & Y \\
05197+3355  & L & 05:23:04.4 & +33:58:27  &   0.0  &   N  & N & N & N \\
05236+3828  & L & 05:27:06.6 & +38:31:18  &  0.0  &   N  & N & N & N \\
05274+3345  & H & 05:30:45.6 & +33:47:52  &  -3.8  &   Y  & -- & -- &  Y \\
05334+3149  & H & 05:36:41.1 & +31:51:14 &  0.0  &   N  & N & N & N \\
05345+3157  & L & 05:37:47.8 & +31:59:24 &   -18.4  &   N  & N & N & Y \\ 
05351+3549\tablefootmark{a}   & H & 05:38:28.1 & +35:51:15  &  0.0  &   N  & -- & -- &  N\tablefootmark{w}  \\
05355+3039  & L & 05:38:47.2 & +30:41:27  &  0.0  &   N  & N & N & N \\
05358+3543\tablefootmark{a}   & H & 05:39:10.4 & +35:45:19  &  0.0  &   Y  & -- & -- & Y\tablefootmark{v} \\  
05373+2349  & L & 05:40:24.4 & +23:50:54  &  2.3  &   N  & N & N &  Y \\
05382+3547  & L & 05:41:37.4 & +35:48:49  &  0.0  &   N  & -- & -- &  N \\
05480+2545  & H & 05:51:10.6 & +25:46:14  &  0.0  &  {\it Y}  & {\it Y} & Y & N \\
05490+2658  & H & 05:52:12.8 & +26:59:32  &  0.0  &   N  & N & N & N \\
05553+1631  & H & 05:58:13.9 & +16:32:00  &  6.1  &   N  & N & N &  Y \\
05554+2013  & H & 05:58:24.5 & +20:13:57  &  0.0  &   N  & N  & N & N \\
05588+2625  & L & 06:01:55.2 & +26:24:59  &  0.0  &   N  & N & N & N \\
06001+3014\tablefootmark{a}   & L & 06:03:22.9 & +30:14:11  &  0.0  &   N  & -- & -- &  Y\tablefootmark{v} \\
06013+3030  & L & 06:04:34.2 & +30:30:39  &  0.0 &   N  & N & N & N \\
06055+2039\tablefootmark{a}   & H & 06:08:32.8 & +20:39:16  &  0.0  &   Y  & -- & -- &  Y\tablefootmark{v} \\
06056+2131  & H & 06:08:40.9 & +21:31:01  &  2.6  &   Y  & N & N & N  \\
06058+2138\tablefootmark{a}   & H & 06:08:54.1 & +21:38:25  &  0.0  &   Y  & -- & -- & Y\tablefootmark{v}   \\
06061+2151  & H & 06:09:07.8 & +21:50:39  &  -0.6  &   Y  & N & N & Y \\
06063+2040  & H & 06:09:21.9 & +20:39:28  &  0.0  &   N  & -- & -- &  N\\
06068+2030  & L & 06:09:51.7 & +20:30:04  &  0.0  &   N  & -- & -- & N \\
06084+1727\tablefootmark{a}   & H & 06:11:24.5 & +17:26:26  &  0.0  &   N  & -- & -- &  N\tablefootmark{w}   \\
06099+1800\tablefootmark{a}  & L & 06:12:53.3 & +17:59:22  &  0.0  &   Y  & -- & -- &  Y\tablefootmark{v} \\ 
06103+1523  & H & 06:13:15.1 & +15:22:36  &  15.6  &   N  & {\it Y\tablefootmark{m}} & N & N \\
06104+1524  & H & 06:13:21.3 & +15:23:57  &  0.0  &   N  & -- & -- &  N \\
06105+1756  & H & 06:13:28.3 & +17:55:29  &  8.0  &   N  & N & N &  Y \\
06114+1745\tablefootmark{a}   & H & 06:14:23.7 & +17:44:36  &  0.0  &   N  & -- & -- & N\tablefootmark{v}   \\
06117+1350\tablefootmark{a}   & H & 06:14:36.6 & +13:49:35  &  0.0  &   Y  & -- & -- & Y\tablefootmark{v}   \\
06155+2319  & H & 06:18:35.2 & +23:18:11  &  0.0  &   N  & -- & -- &  N \\
06158+1517\tablefootmark{a}   & L & 06:18:44.8 & +15:16:43  &  0.0  &   N  & -- & -- & N\tablefootmark{w}   \\ 
06291+0421  & L & 06:31:48.1 & +04:19:31  &  0.0  &   N  & -- & -- & Y \\
06299+1011  & L & 06:32:41.3 & +10:09:34  &  0.0  &   N  & -- & -- & N \\
06303+1021  & L & 06:33:04.4 & +10:19:21  &  0.0  &   N  & -- & -- &  N\\
06308+0402  & L & 06:33:31.1 & +04:00:07  &  0.0  &   N  & -- & -- &  N \\
06314+0427  & L & 06:34:05.5 & +04:24:54  &  0.0  &   N  & N & N & N \\
06382+0939  & L & 06:41:02.7 & +09:36:10  &  5.2  &   N  & N & N & N \\
06568-0411  & L & 06:59:22.4 & -04:15:45  &  0.0  &   N  & -- & -- & N \\
06584-0852  & L & 07:00:51.5 & -08:56:29  &  41.3 &   N  & N & N &  Y \\
07029-1215\tablefootmark{a}   & L & 07:05:16.9 & -12:20:02  &  0.0  &   N  & -- & -- & N\tablefootmark{h}   \\
07069-1045  & H & 07:09:20.5 & -10:50:25  &  0.0  &   N  & N & N & N \\
07284-1511\tablefootmark{a}   & H & 07:30:42.5 & -15:17:40  &  0.0  &   N  & -- & -- &  N\tablefootmark{b}  \\
07295-1915  & L & 07:31:47.6 & -19:21:48  &  0.0  &   N  & -- & -- & N \\
07333-1838\tablefootmark{a}   & H & 07:35:34.3 & -18:45:32  &  0.0  &   N  & -- & -- & N\tablefootmark{v}   \\
07429-2523\tablefootmark{a}   & L & 07:45:01.4 & -25:31:10  &  0.0  &   N  & -- & -- & N\tablefootmark{w}   \\ 
17417-2851  & H & 17:44:53.4 & -28:52:20  &  -5.6  &   N  & -- & -- & N \\
17450-2742  & L & 17:48:09.3 & -27:43:21 &   -16.9  &   Y  & -- & -- & N \\
17456-2850  & H & 17:48:47.1 & -28:51:49  &  0.0  &   N  & -- & -- & N \\
17495-2624  & H & 17:52:41.9 & -26:25:32  &  17.3  &   N  & -- & -- & N \\
17504-2519  & H & 17:53:35.1 & -25:19:56  &  12.4  &   N  & -- & -- & N \\
17527-2439  & H & 17:55:49.1 & -24:40:20  &  13.2  &   N  & N & N &  Y \\
17571-2328  & H & 18:00:14.2 & -23:28:57  &  0.0  &   {\it Y}  & -- & -- & N \\
17580-2215  & L & 18:01:02.2 & -22:15:45  &  0.0  &   N  & -- & -- & N \\
17582-2234  & L & 18:01:18.0 & -22:34:57  &  0.0  &   N  & -- & -- & Y  \\
17589-2312  & H & 18:01:57.1 & -23:12:36  &  0.0  &   Y  & -- & -- & N \\
18014-2428  & L & 18:04:29.6 & -24:28:47  &  12.5  &   N  & N & N & N \\
18018-2426  & L & 18:04:53.9 & -24:26:41  &  10.5  &   N  & Y & Y &  N \\
18024-2119  & L & 18:05:25.4 & -21:19:41  &  0.5  &   {\it Y } & {\it Y} & N & Y \\
18024-2231  & L & 18:05:30.6 & -22:31:35  &  16.2  &   N  & -- & -- & N \\
18026-2145  & L & 18:05:37.9 & -21:45:06  &  0.0  &   N  & -- & -- & N \\
18027-2202  & H & 18:05:46.2 & -22:01:57  &  0.0  &   N  & -- & -- & N \\
18039-2052  & L & 18:06:56.7 & -20:51:51  &  0.0  &   N  & -- & -- & N \\
18043-2153  & L & 18:07:23.2 & -21:53:00  &  0.0  &   N  & -- & -- &  N \\
18048-2019  & H & 18:07:51.5 & -20:18:36  &  49.1  &   Y  & N & N & Y \\ 
18054-1818  & L & 18:08:24.0 & -18:17:35  &  0.0  &   N  & -- & -- &  N \\
18078-1928  & H & 18:10:48.7 & -19:27:42  &  0.0  &   N  & -- & -- & N \\
18079-1756  & H & 18:10:50.3 & -17:55:52  &  0.0  &   N  & -- & -- & N \\
18081-2138  & L & 18:11:10.5 & -21:38:15  &  0.0  &   N  & -- & -- &  N \\
18089-1732  & H & 18:11:51.3 & -17:31:29  &  32.9  &   Y  & Y &  Y & Y \\ 
18102-1800  & H & 18:13:12.1 & -17:59:34  &  0.0  &   Y  & -- & -- & N \\
18115-1701  & L & 18:14:25.6 & -17:00:53  &  0.0  &   N  & -- & -- & N \\
18122-1751  & L & 18:15:12.2 & -17:49:60  &  0.0  &   N  & -- & -- & N \\
18123-1203  & L & 18:15:07.3 & -12:02:42  &  0.0  &   N  & -- & -- &  N \\
18131-1606  & L & 18:16:01.9 & -16:05:05  &  0.0  &   N  & -- & -- & N \\
18132-1638  & L & 18:16:08.8 & -16:37:07  &  0.0  &   N  & -- & -- & N \\
18134-1942  & H & 18:16:22.2 & -19:41:20  &  10.5  &   Y  & N & N & Y \\ 
18136-1347  & L & 18:16:28.6 & -13:46:33  &  0.0  &   N  & -- & -- & N \\
18144-1723  & L & 18:17:24.4 & -17:22:13  &  47.3  &   Y  & Y & Y &  Y \\
18145-1557  & H & 18:17:26.7 & -15:56:19  &  0.0  &   N  & -- & -- & N \\
18151-1208  & H & 18:17:57.1 & -12:07:22  &  32.8  &   Y  & Y & N &  N \\
18153-1651  & H & 18:18:15.9 & -16:50:37  &  0.0  &   N  & -- & -- & N \\
18156-1343  & L & 18:18:28.6 & -13:42:22  &  0.0  &   N  & -- & -- &  N \\
18159-1550  & H & 18:18:47.6 & -15:48:54  &  59.7  &   N  & -- & -- & N \\ 
18159-1648  & H & 18:18:53.0 & -16:47:39  &  22.1 &   Y  & Y & Y & N \\
18160-1641  & L & 18:18:57.3 & -16:40:15  &  0.0  &   N  & -- & -- &  N \\
18162-1612  & L & 18:19:07.5 & -16:11:21  &  61.8  &   N  & Y & N & N \\
18167-1614  & L & 18:19:37.7 & -16:13:11 &  0.0  &   N  & -- & -- & N \\
18171-1548  & H & 18:20:04.6 & -15:46:47  &  0.0  &   N  & -- & -- & N \\
18181-1534  & L & 18:20:58.9 & -15:33:10  &  0.0 &  Y  & -- & -- & N \\
18195-1407  & L & 18:22:24.4 & -14:05:33  &  0.0 &  N  & -- & -- & N \\
18197-1351  & H & 18:22:36.7 & -13:50:11 &  0.0 &  N  & -- & -- & N \\
18198-1429  & H & 18:22:41.7 & -14:27:35  &  0.0 &  N  & -- & -- & N \\
18212-1320  & H & 18:24:04.7 & -13:19:16  &  0.0 &  N  & -- & -- & N \\
18224-1228  & L & 18:25:12.6 & -12:27:09  &  0.0 &  Y  & -- & -- &N \\
18230-1248  & L & 18:25:51.7 & -12:46:15  &  0.0 &  N  & -- & -- & N  \\
18236-1205  & H & 18:26:24.3 & -12:03:47  &  26.2 &  Y  & -- & -- & Y \\
18247-1147  & H & 18:27:31.1 & -11:45:56  &  55.9 &  {\it Y}  & -- & -- & N  \\
18248-1218  & L & 18:27:37.9 & -12:16:27  &  0.0 &  N  & -- & -- & N \\
18256-0742  & L & 18:28:20.5 & -07:40:22  &  36.7 &  N  & N & N & N \\
18258-0737  & H & 18:28:34.1 & -07:35:31  &  37.9 &  N  & -- & -- & N \\
18265-1517  & H & 18:29:24.8 & -15:15:49  &  0.0 &  Y  & -- & -- & N \\
18278-1009  & L & 18:30:35.2 & -10:07:12  &  93.7 &  Y  & N & N & N \\
18288-0158  & L & 18:31:26.6 & -01:56:35  &  5.8 &  N  & N & N & N \\
18310-0825  & H & 18:33:47.2 & -08:23:34  &  0.0 &  N  & -- & -- & N \\
18311-0701  & L & 18:33:52.8 & -06:58:44  &  0.0 &  N  & -- & -- & N \\
18314-0820  & H & 18:34:08.9 & -08:17:52  &  0.0 &  N  & -- & -- & N \\
18314-0517  & L & 18:34:05.9 & -05:14:49  &  0.0 &  N  & -- & -- &  N \\
18316-0602  & H & 18:34:19.8 & -05:59:44  &  42.2 &  Y  & {\it Y} & Y$^{\rm m}$ & Y \\
18317-0845  & H & 18:34:29.9 & -08:43:22  &  0.0 &  Y  & -- & -- & N \\
18317-0513  & H & 18:34:25.9 & -05:10:59  &  42.2 &  N  & -- & -- & N \\
18318-0741  & L & 18:34:30.8 & -07:38:54  &  0.0 &  N  & -- & -- & N \\
18355-0550  & L & 18:38:13.6 & -05:48:17  &  0.0 &  N  & -- & -- & N \\
18358-0647  & H & 18:38:31.3 & -06:44:29  &  0.0 &  N  & -- & -- &  N \\
18360-0537  & H & 18:38:40.3 & -05:35:06  &   102.3 &  N  & {\it Y} & Y & Y \\
18363-0554  & L & 18:39:03.7 & -05:52:15  &  65.2 &  N  & N & N & N \\
18372-0541  & H & 18:39:56.0 & -05:38:49  &  22.8 &  Y  & -- & -- & N \\
18396-0431  & L & 18:42:18.8 & -04:28:37  &  97.3 &  N  & {\it Y$^{\rm m}$} & Y & N \\
18408-0353  & H & 18:43:28.8 & -03:50:20  &  0.0 &  {\it Y}  & -- & -- & N \\
18408-0348  & H & 18:43:31.1 & -03:44:57  &  0.0 &  N  & -- & -- & N \\
18411-0312  & L & 18:43:46.4 & -03:09:54  &  0.0 &  N  & -- & -- & N  \\
18424-0329  & L & 18:45:03.3 & -03:26:49  &  47.9 &  N  & N & N & N \\
18441-0134  & L & 18:46:44.2 & -01:30:55  &  0.0 &  Y  & -- & -- & N \\
18445-0222  & H & 18:47:10.7 & -02:19:06  &  86.9 &  N  & -- & -- & N \\
18463-0052  & L & 18:48:52.8 & -00:49:02  &  0.0 &  N  & -- & -- & N \\
18488+0000  & H & 18:51:24.8 & +00:04:19  &  83.2 &  Y  & -- & -- & N \\
18507+0121  & H & 18:53:17.4 & +01:24:55  &  57.1 &  Y  & {\it Y} & N & Y \\
18511+0146  & L & 18:53:38.1 & +01:50:27  &  56.8 & {\it  Y}  & -- & -- & N \\
18517+0437  & H & 18:54:13.8 & +04:41:32  &  43.7 &  Y  & {\it Y} & N & Y \\
18527+0301  & L & 18:55:16.5 & +03:05:07 &  76.0 &  Y  & N  & N & N \\
18532+0047  & H & 18:55:50.8 & +00:51:22  &  58.6 &  {\it Y}  & -- & -- & N \\
18537+0145  & L & 18:56:15.0 & +01:49:03  &  0.0 &  N  & -- & -- & N \\
18544+0112  & L & 18:56:59.8 & +01:16:20  &  0.0 &  N  & -- & -- & N \\
18551+0302  & H & 18:57:42.1 & +03:06:04  &  57.5 &  N  & -- & -- & N \\
18565+0349  & L & 18:59:03.4 & +03:53:22  &  91.6 &  N  & N & N & N \\
18566+0408  & H & 18:59:09.8 & +04:12:14  &  85.1 &  Y  & -- & -- & N \\
18567+0700  & L & 18:59:13.6 & +07:04:47  &  29.4 &  N  & N & N & N  \\
18571+0326  & L & 18:59:40.2 & +03:30:29 &  0.0 &  N  & -- & -- & N \\
18571+0349  & L & 18:59:40.1 & +03:53:35  &  56.2 &  N  & -- & -- & N \\
18586+0106  & L & 19:01:10.6 & +01:11:16  &  37.9 &  N  & N & N &  N \\
18596+0536  & H & 19:02:06.5 & +05:40:33  &  0.0 &  N  & -- & -- & N \\
19001+0402  & H & 19:02:36.2 & +04:06:58  &  50.7 &  N  & -- & -- & N \\
19002+0454  & H & 19:02:42.0 & +04:58:49  &  69.2 &  N  & -- & -- & N \\
19012+0505  & L & 19:03:43.5 & +05:09:49  &  40.4 &  Y  & -- & -- & N \\
19023+0538  & L & 19:04:49.0 & +05:42:40  &  0.0 &  N  & -- & -- & N \\
19043+0726  & H & 19:06:47.6 & +07:31:38  &  58.9 &  N  & {\it Y} & N & N \\
19045+0518  & H & 19:06:59.3 & +05:22:57 &  53.6 &  N  & -- & -- & N \\
19045+0813  & H & 19:06:59.8 & +08:18:43  &  0.0 & N  & -- & -- & Y \\
19077+0839  & L & 19:10:09.6 & +08:44:11  &  0.0 &  N  & -- & -- & N \\
19088+0902  & H  & 19:11:15.9 & +09:07:27 &  59.6 &  N  & {\it Y} & N & Y \\
19092+0841  & L & 19:11:37.4 & +08:46:30 &  58.0 &  Y  & Y & N & Y \\
19094+0944  & H & 19:11:52.0 & +09:49:46 &  65.3 &  N  & -- & -- & N \\
19118+0945  & L & 19:14:14.7 & +09:50:40  &  0.0 &  N  & -- & -- & N \\
19120+1148  & L & 19:14:22.1 & +11:53:39 &  0.0 &  {\it Y}  & -- & -- & N \\
19124+1106  & H & 19:14:45.3 & +11:11:57  &  0.0 &  N  & -- & -- & N \\ 
19139+1137  & L & 19:16:18.7 & +11:42:46  &  0.0 &  N  & -- & -- & N \\
19183+1556  & L & 19:20:37.6 & +16:02:31  &  0.0 &  N  & -- & -- & N \\
19186+1440  & L & 19:20:56.9 & +14:46:40  &  0.0 &  Y  & -- & -- & N \\
19198+1423  & H & 19:22:07.7 & +14:29:20 &  58.9 &  N  & -- & -- & N \\
19202+1412  & L & 19:22:31.5 & +14:18:30  &  0.0 &  N  & -- & -- & N \\
19213+1723  & H & 19:23:37.0 & +17:28:59 &  41.7 &  N  & N & N & Y \\
19214+1458  & H & 19:23:46.1 & +15:04:51  &  0.0 &  N  & -- & -- & N \\
19230+1341  & H & 19:25:22.1 & +13:47:19  &  0.0 &  Y  & -- & -- & N \\ 
19282+1814  & H & 19:30:28.1 & +18:20:53 &  24.1 &  Y  & N & N & N \\
19282+1742  & H & 19:30:30.4 & +17:48:30  &  0.0 &  N  & -- & -- & N \\
19294+1649  & L & 19:31:44.8 & +16:55:57  &  0.0 &  N  & -- & -- &  N \\
19295+1637  & L & 19:31:50.9 & +16:43:29  &  0.0 &  N  & -- & -- & N \\
19325+1925  & H  & 19:34:45.9 & +19:31:41 &  0.0 &  N  & -- & -- & N \\
19332+2028  & L & 19:35:25.4 & +20:34:56  &  0.0 &  N  & -- & -- & N  \\
19343+2026  & L & 19:36:30.2 & +20:33:08  &  0.0 &  N  & -- & -- & N \\
19368+2239  & H & 19:38:58.1 & +22:46:32  &  36.4 &  N  & -- & -- &  N \\
19374+2352  & H & 19:39:33.2 & +23:59:55  &  36.9 &  N  & N  & N  & Y \\
19383+2711  & H & 19:40:22.1 & +27:18:34  &  0.0 &  N  & -- & -- &  N \\
19388+2357  & H & 19:40:59.3 & +24:04:39  &  34.6 &  Y  & {\it Y} & N & Y \\
19423+2541  & H & 19:44:23.4 & +25:48:40  &  0.0 &  N  & -- & -- &  N \\
19458+2442  & L & 19:47:59.7 & +24:50:19  &  0.0 &  N  & -- & -- & N \\
19462+2759  & H & 19:48:14.0 & +28:07:23  &  0.0 &  N  & -- & -- & N \\
19499+2613  & H & 19:52:01.4 & +26:21:10  &  0.0 &  N  & -- & -- & N \\
19542+3004  & L & 19:56:13.2 & +30:12:56  &  0.0 &  N  & -- & -- & N \\
19560+3135  & H & 19:58:03.3 & +31:44:06  &  0.0 &  N  & -- & -- &  N \\
19592+3302  & H & 20:01:09.8 & +33:11:08  &  0.0 &  N  & -- & -- & N \\
20028+2903  & L & 20:04:53.3 & +29:11:37  &  0.0 &  N  & -- & -- & N \\
20050+2720  & H & 20:07:06.7 & +27:28:53  &  6.4 &  N  & {\it Y} & N & Y  \\
20051+3435  & H & 20:07:03.8 & +34:44:34 &  0.0 &  N  & -- & -- & N \\
20056+3350  & H & 20:07:31.5 & +33:59:39  &  9.4 &  N  & N & N & Y \\
20062+3550  & H & 20:08:09.7 & +35:59:20  &  0.6 &  Y  & {\it Y} & N & Y \\
20081+2720  & H & 20:10:11.4 & +27:29:06  &  0.0 &  N  & -- & -- & N \\
20099+3640  & L & 20:11:46.4 & +36:49:37 &   -36.4 &  N  & N & N &N  \\
20103+3633  & H & 20:12:13.9 & +36:42:58  &  0.0 &  N  & -- & -- & N \\
20106+3545  & L & 20:12:31.3 & +35:54:46  &  7.8 &  N  & N & N & N \\
20126+4104  & H & 20:14:26.0 & +41:13:32  &  -3.9 &  Y  & Y & Y &  Y \\
20149+3440  & L & 20:16:50.4 & +34:49:22  &  0.0 &  N  & -- & -- & N \\
20153+3850\tablefootmark{a}   & L & 20:17:07.6 & +38:59:25  &  0.0 &  N  & -- & -- & N\tablefootmark{b} \\
20160+3636\tablefootmark{a}   & H & 20:17:56.1 & +36:45:33  &  0.0 &  N  & -- & -- &  N\tablefootmark{b} \\
20173+3714  & L & 20:19:10.4 & +37:23:31  &  0.0 &  N  & -- & -- & N \\
20180+3558  & L & 20:19:58.0 & +36:07:37  &  0.0 &  N  & -- & -- & N \\
20188+3928  & H & 20:20:39.3 & +39:37:52  &  1.5 &  N  & {\it Y} & N & Y \\
20190+4102  & L & 20:20:47.9 & +41:12:08  &  0.0 &  N  & -- & -- &N  \\
20205+3948  & H & 20:22:21.8 & +39:58:04  &  0.0 &  N  & -- & -- & N \\
20216+4107  & H & 20:23:23.8 & +41:17:40  &  0.0 &  N  & -- & -- & N \\
20217+3947  & L & 20:23:31.7 & +39:57:23  &  -0.9 &  N  & N & N & N \\
20220+3728  & H & 20:23:55.6 & +37:38:10 &  -2.7 &  N  & -- & -- & N \\
20222+3541  & L & 20:24:10.9 & +35:51:37  &  0.0 &  N  & -- & -- & N \\
20227+4154  & H & 20:24:31.4 & +42:04:17  &  5.9 &  N  & N & N & Y \\
20228+4215\tablefootmark{a}   & L & 20:24:34.4 & +42:25:01  &  0.0 &  N  & -- & -- & N\tablefootmark{b}  \\
20243+3853  & H & 20:26:10.8 & +39:03:29  &  0.0 &  N  & -- & -- & N \\
20261+3825  & L & 20:28:01.9 & +38:35:50  &  0.0 &  N  & -- & -- &  N \\
20264+4042  & H & 20:28:12.3 & +40:52:28  &  0.0 &  N  & -- & -- & N \\
20277+3851  & H & 20:29:36.7 & +39:01:17  &  0.0 &  N  & -- & -- & N \\
20278+3521  & L & 20:29:46.9 & +35:31:39 &  -4.5 &  N  & N & N & N \\
20281+4038  & H & 20:29:54.7 & +40:48:52  &  0.0 &  N  & -- & -- & N \\
20286+4105  & H & 20:30:27.9 & +41:15:48 &  -3.8 &  N  & -- & -- & Y \\
20293+4007  & L & 20:31:07.9 & +40:17:23  &  0.0 &  N  & -- & -- & N \\
20306+4005  & H & 20:32:28.7 & +40:16:05  &  0.0 &  N  & -- & -- & N \\
20306+3749  & H & 20:32:34.4 & +37:59:35  &  0.0 &  N  & -- & -- & N \\
20319+3958  & H & 20:33:49.4 & +40:08:44  &  0.0 &  N  & -- & -- &  N \\
20321+4112  & H & 20:33:55.6 & +41:22:43  &  0.0 &  N  & -- & -- & N \\
20327+4120  & L & 20:34:31.1 & +41:30:44  &  0.0 &  N  & -- & -- & N \\
20332+4124  & H & 20:35:00.5 & +41:34:48  &  0.0 &  N  & -- & -- & N \\
20333+4102  & L & 20:35:09.5 & +41:13:18  &  8.4 &  N  & -- & -- & N \\
20337+4104  & L & 20:35:34.7 & +41:15:21  &  0.0 &  N  & -- & -- &  N \\
20343+4129  & H & 20:36:07.1 & +41:40:01 &  0.0 &  N  & -- & -- & N \\
20350+4126  & H & 20:36:52.5 & +41:36:33  &  0.0 &  Y  & -- & -- &  N \\
20406+4555  & L & 20:42:21.6 & +46:05:53  &  0.0 &  N  & -- & -- & N \\
20444+4629  & H & 20:46:08.2 & +46:40:41 &  -4.1 &  N  & N & N & N \\
20446+4613  & L & 20:46:17.3 & +46:24:38 &  0.0 & N  & -- & -- & N \\
21020+4939\tablefootmark{a}   & L &  21:03:42.2 & +49:51:53  &  0.0 &  N  & -- & -- & N\tablefootmark{w}   \\   
21036+4927\tablefootmark{a}   & H & 21:05:15.6 & +49:40:01  &  0.0 &  N  & -- & -- & N\tablefootmark{w}    \\  
21046+5110  & L & 21:06:16.3 & +51:22:13  &  0.0 &  N  & -- & -- & N \\
21078+5211  & H & 21:09:25.2 & +52:23:44 &  -6.1 &  N  & N & N & Y \\
21080+4758  & L & 21:09:46.5 & +48:10:59 &  0.0 &  N  & -- & -- & N \\
21202+5157  & H & 21:21:53.2 & +52:10:44  &  0.0 &  N  & -- & -- & N \\
21270+5423\tablefootmark{a}   & L & 21:28:42.0 & +54:36:51  &  0.0 &  N  & -- & -- &   N\tablefootmark{w}   \\
21306+5540\tablefootmark{a}   & H & 21:32:11.6 & +55:53:24  &  0.0 & {\it Y}  & -- & -- &  Y\tablefootmark{v}  \\
21307+5049  & L & 21:32:31.5 & +51:02:22 &   -46.7&  N  & N & N & Y \\
21336+5333  & H & 21:35:21.0 & +53:47:12  &  0.0 &  N  & -- & -- & N \\
21391+5802  & H & 21:40:42.3 & +58:16:10  &  0.4 &  N  & Y & Y & Y \\
21418+6552\tablefootmark{a}  & H & 21:43:02.3 & +66:06:29  &  0.0 &  N  & -- & -- &  Y\tablefootmark{v}   \\
21418+5403\tablefootmark{a}  & H & 21:43:29.8 & +54:16:56  &  0.0 &  N  & -- & -- &  N\tablefootmark{w}   \\
21519+5613  & H & 21:53:39.2 & +56:27:46 &   -63.2 & N  & -- & -- & N \\
21526+5728  & H & 21:54:18.4 & +57:42:51  &  0.0 &  N  & -- & -- & N \\
21548+5747  & L & 21:56:29.9 & +58:01:35  &  0.0 &  N  & -- & -- & N \\
22147+5948  & L & 22:16:28.6 & +60:03:49  &  0.0 &  N  & -- & -- & N \\
22172+5549  & L & 22:19:09.0 & +56:04:45 &   -43.8 &  N  & N & N & N \\
22187+5559  & L & 22:20:34.9 & +56:14:40  &  0.0 &  {\it Y}  & -- & -- & N \\
22198+6336  & H & 22:21:27.6 & +63:51:42 &   -11.1 &  N  & N & N & Y \\
22267+6244  & H & 22:28:29.3 & +62:59:44 &  -1.5 &  N  & N & N & N  \\
22272+6358  & H & 22:28:52.2 & +64:13:43 &  -9.9 &  Y  & N & N & N \\
22305+5803  & H & 22:32:24.3 & +58:18:58 &   -52.1 &  N  & -- & -- & Y  \\
22344+5909  & L & 22:36:20.6 & +59:24:57  &  0.0 &  N  & -- & -- & N \\
22457+5751  & L & 22:47:46.5 & +58:07:19  &  0.0 & N  & -- & -- & N \\
22475+5939\tablefootmark{a}  & H & 22:49:29.5 & +59:54:57  &  0.0 &  N  & -- & -- &  Y\tablefootmark{v} \\    
22506+5944  & H & 22:52:38.6 & +60:00:56 &   -51.5 & {\it Y}  & {\it Y} & Y$^{\rm m}$ & Y \\
22539+5758\tablefootmark{a}  & H & 22:56:00.0 & +58:14:46  &  0.0 &  N  & -- & -- &  N\tablefootmark{v}     \\
22542+5815\tablefootmark{a}  & L & 22:56:17.0 & +58:31:13  &  0.0 &  N  & -- & -- &   N\tablefootmark{v}  \\
22543+6145\tablefootmark{a}  & H & 22:56:19.1 & +62:01:57  &  0.0 &  Y  & -- & -- &   Y\tablefootmark{v}  \\
22551+6221  & H & 22:57:05.2 & +62:37:44  &  0.0 &  N  & -- & -- & N  \\
22566+5830\tablefootmark{a}  & H & 22:58:41.3 & +58:46:57  &  0.0 &  N  & -- & -- &   Y\tablefootmark{v}  \\
22570+5912  & H & 22:59:06.4 & +59:28:28 &  0.0 &  N  & -- & -- & N \\
23026+5948  & L & 23:04:45.7 & +60:04:35 &   -51.1 &  N & N & N & N \\
23042+6000  & L & 23:06:21.2 & +60:16:16  &  0.0 &  N  & N & N &  N \\
23134+6131  & L & 23:15:34.5 & +61:47:41  &  0.0 &  N  & N & N & N \\
23140+6121  & L & 23:16:11.7 & +61:37:45 &   -51.5 &  N  & N & N & N \\
23146+5954  & L & 23:16:48.9 & +60:10:46 &  0.0 &  N  & -- & -- & N \\
23152+6034  & L & 23:17:25.8 & +60:50:45  &  0.0 &  N  & -- & -- & N \\
23314+6033  & L & 23:33:44.4 & +60:50:30 &   -45.4 &  N  & N & N & N \\
23330+6437  & L & 23:35:23.7 & +64:54:31  &  0.0 &  N  & -- & -- & N \\
23385+6053  & L & 23:40:53.2 & +61:10:21 &   -50.0 &  N  & Y & N & Y \\
23448+6010  & L & 23:47:20.2 & +60:27:21  &  0.0  & N & -- & -- & N \\
23504+6012\tablefootmark{a}  & L & 23:52:58.2 & +60:28:45  &  0.0 &  N  & -- & -- & N\tablefootmark{w}   \\    
23507+6230  & L & 23:53:12.9 & +62:47:00  &  0.0 &  N  & -- & -- & N \\
23545+6508  & H & 23:57:05.2 & +65:25:11 &   -18.4 &  N  & -- & -- & N \\
\hline
\end{longtable}
\tablefoot{
\tablefoottext{p}{Observed by Palla et al.~(\citeyear{palla}) with the Medicina 32-m telescope, except when differently specified;}
\tablefoottext{a}{Rejected by Palla et al.~(\citeyear{palla}) because associated with radio continuum;}
\tablefoottext{w}{Observed by Wouterloot et al.~(\citeyear{wouterloot});}
\tablefoottext{v}{Observed by Valdettaro et al.~(\citeyear{valdettaro});}
\tablefoottext{m}{Marginal detection at $\sim 3 \sigma$;}
\tablefoottext{h}{Observed by Henning et al.~(\citeyear{henning});}
\tablefoottext{b}{Observed with the Medicina 32-m telescope (unpublished data).}
}
}  

\addtocounter{table}{2}

\longtab{2}{
\begin{longtable}{cccccc}
\caption{\label{par_classII_high} Parameters of the 6 GHz \METH\ maser (Class II) detected in \high\ sources with the Effelsberg 100-m
telescope.} \\
\hline \hline
Source & \Vlsr\ \tablefootmark{a} & $\int F_\nu {\rm d}V$  &    $V$     &      $\Delta V$    &     $F_{\rm peak}$  \\
        & (km s$^{-1}$)  &  (Jy km s$^{-1}$)  & (km s$^{-1}$)  &  (km s$^{-1}$)  &  (Jy)  \\
\hline
\endfirsthead
\caption{continued.} \\
\hline\hline
Source & \Vlsr\ \tablefootmark{a} & $\int F_\nu {\rm d}V$  &    $V$     &      $\Delta V$    &     $F_{\rm peak}$  \\
        & (km s$^{-1}$)  &  (Jy km s$^{-1}$)  & (km s$^{-1}$)  &  (km s$^{-1}$)  &  (Jy)  \\
\hline
\endhead
\hline
\endfoot
00338+6312* &   -17.5$^{\rm f}$  &    6.4$\pm$0.1        &  -23.030$\pm$0.001 &     0.287$\pm$0.003 &        21.0          \\    
                         &           &     2.8$\pm$0.2  &          -22.64$\pm$0.02 &     0.74$\pm$0.04  &         3.62           \\
05274+3345  &   -3.8  &      2.40$\pm$0.04 &        0.324$\pm$0.002 &    0.331$\pm$0.006  &     6.81           \\  
    &      &    10.07$\pm$0.04 &         1.244$\pm$0.003  &   0.584$\pm$0.004   &   16.2           \\
    &       &    16.59$\pm$0.04        & 1.609$\pm$0.0005   &  0.349$\pm$0.003   &   44.6           \\ 
   &    &      1.15$\pm$0.03 &         2.140$\pm$0.003 &    0.202$\pm$0.005   &   5.35           \\
05358+3543* & -17.6\tablefootmark{d}  &  18$\pm$1        &          -13.8$\pm$0.1 &              0.5$\pm$0.1        &  34.4 \\
   &     & 15$\pm$1         &         -13.2$\pm$0.1    &              0.3$\pm$0.1        &  45.7  \\
   &     &  19$\pm$1         &         -12.8$\pm$0.1   &        0.3$\pm$0.1  &            51.8     \\
   &     &  1.5$\pm$1         &         -11.5$\pm$0.1   &          3.0$\pm$0.1  &   0.47  \\
05480+2545    &  --9.3\tablefootmark{b}   &     0.98$\pm$0.03 &         -15.250$\pm$0.004  &   0.238$\pm$0.009 &        3.88           \\
    &      &    3.09$\pm$0.04 &        -14.730$\pm$0.003   &  0.426$\pm$0.007  &    6.81           \\ 
     &       &   0.27$\pm$0.03 &         -13.99$\pm$0.02  &   0.30$\pm$0.05 &     0.84   \\
     &        & 6.15$\pm$0.04 &        -4.8940$\pm$0.0001 &    0.340$\pm$0.003   &   17.0          \\
 06055+2039*  &  8.8\tablefootmark{c}    &  0.2$\pm$0.02   &     3.9$\pm$0.02  &        0.38$\pm$0.06 &     0.5 \\ 
                                  &          &   0.78$\pm$0.02  &      5.496$\pm$0.006  &        0.47$\pm$0.02   &      1.57 \\  
06056+2131    &    2.6    &    3.7$\pm$0.2 &        8.8$\pm$0.1        &  0.3$\pm$0.1          &    12.7          \\
  &         &   2.2$\pm$0.2 &        9.6$\pm$0.1        &  0.3$\pm$0.1           &   7.26          \\ 
   &        &   0.7$\pm$0.2 &        10.6$\pm$0.1        &  1.1$\pm$0.1          &   0.605          \\ 
   06058+2138* & 3.1\tablefootmark{c}  &      4.1$\pm$0.2 &   8.470$\pm$0.009 &     0.43$\pm$0.03 &         9.10 \\
                               &        &  5.0$\pm$0.1 &          9.446$\pm$0.001 &     0.386$\pm$0.008 &        12.2 \\
                              &        &  373.0$\pm$0.2 &          10.590$\pm$0.001 &     0.784$\pm$0.001 &        447 \\
                              &        &  17.7$\pm$0.1   &          11.45$\pm$0.01 &     0.212$\pm$0.002 &        78.7  \\
06061+2151   &    -0.6    &    0.51$\pm$0.02 &         -5.564$\pm$0.003          & 0.232$\pm$0.008    &  2.08        \\
    &         &   0.17$\pm$0.03 &         -5.05$\pm$0.06         &  0.7$\pm$       0.2         &    0.227          \\
   &        &   0.30$\pm$0.02 &          10.63$\pm$0.03         &  0.74$\pm$0.07 &    0.380        \\
  06117+1350*  & 17.9\tablefootmark{c}    &  2.2$\pm$0.8   &   14.6$\pm$0.1        &  0.2$\pm$0.1         &    9.5 \\
                                &           &  25.0$\pm$0.8  &     15.1$\pm$0.1        &  0.5$\pm$0.1         &    47.8 \\
                                &          &  4.6$\pm$0.8    &  15.7$\pm$0.1  &         0.5$\pm$0.1        &     8.2 \\
17571-2328     &   --      &  0.85$\pm$0.01 &          143.80$\pm$0.03          & 1.16$\pm$0.05   &   0.686        \\
    &       &    0.29$\pm$0.03 &          147.30$\pm$0.02         &  0.37$\pm$0.05  &   0.731         \\        
17589-2312    &    21.1\tablefootmark{c}     &  1.5$\pm$0.5         &     13.1$\pm$0.1        &  1.0$\pm$0.1         &     1.44          \\        
     &        &   7.1$\pm$0.5         &14.7$\pm$0.1         &  0.4$\pm$0.1          &    19.1          \\        
     &         &  9.6$\pm$0.5         &15.2$\pm$0.1         & 0.2$\pm$0.1         &     39.3          \\        
     &       &    54.7$\pm$0.5 &        16.0$\pm$0.1 &          0.8$\pm$0.1 &              62.5        \\        
     &       &   2.21$\pm$0.06 &         17.430$\pm$0.006          & 0.48$\pm$0.02  &    4.30  \\        
     &       &   2.59$\pm$0.04 &         18.7000$\pm$0.0008          & 0.239$\pm$0.005  &    10.2        \\        
     &       &   3.80$\pm$        0.05 &         19.540$\pm$0.003 &          0.436$\pm$0.008  &    8.19  \\        
     &       & 1.08$\pm$0.07        &  24.60$\pm$0.04        &   1.19$\pm$0.06   &  0.847        \\        
     &      &    15.60$\pm$0.07 &         25.4500$\pm$0.0008 &          0.569$\pm$0.002 &      25.8 \\
    &      &    26.07        $\pm$0.05        & 26.4800$\pm$0.0009        &  0.937$\pm$0.002   &   26.1 \\
     &     &    10.0$\pm$          0.4 &        28.07$\pm$0.02 &           1.27$\pm$0.04  &    7.37          \\        
    &      &    7.8        $\pm$0.2 &           28.630$\pm$0.003 &          0.422$\pm$0.007  &    17.2 \\        
18048-2019    &  49.1   &     15.3$\pm$0.6 &         41.40$\pm$       0.110  &           0.6142$\pm$        0.110  &               23.3         \\
     &     &    17.7$\pm$0.6 &         42.1$\pm$0.1 &           0.6$\pm$0.1 &               27.3         \\
     &      &    5.9$\pm$0.6 &         43.2$\pm$0.1 &           0.4$\pm$0.1 &               15.5         \\
     &       &    0.7$\pm$0.6 &         44.2$\pm$0.1 &           1.0$\pm$0.1 &              0.629         \\
     &       &   4.1$\pm$0.4  &         45.8$\pm$0.1 &            1.9$\pm$0.1 &               2.03         \\
    &       &   5.6$\pm$0.4  &         46.8$\pm$0.1 &           0.8$\pm$0.1 &               6.67         \\
    &       &   5.2$\pm$0.4  &         47.1$\pm$0.1 &           0.3$\pm$0.1 &               14.6          \\
    &        &  1.4$\pm$0.4  &         48.1$\pm$0.1 &           0.7$\pm$0.1 &               1.82          \\
   &        &  1.0$\pm$0.4  &         50.4$\pm$0.1 &           1.0$\pm$0.1 &               1.00          \\ 
18089-1732    &  32.9      &  2.8$\pm$0.1  &         29.59$\pm$0.08 &     3.4$\pm$0.2 &              0.774        \\
    &        &  0.13$\pm$0.01 &          29.210$\pm$0.004 &    0.111$\pm$0.003   &   1.07         \\
    &        &  0.94$\pm$0.03 &          30.210$\pm$0.003  &  0.270$\pm$0.009    &  3.26         \\
    &        &  3.3$\pm$0.7 &        31.9$\pm$0.1        &  0.7$\pm$0.1        &      4.50         \\
    &        &  4.7$\pm$0.7 &        32.8$\pm$0.1         & 0.5$\pm$0.1         &     9.18         \\
    &       &  24.7$\pm$0.7 &         33.7$\pm$0.1         & 0.5$\pm$0.1        &      50.3         \\
    &        &  3.8$\pm$0.7 &        34.7$\pm$0.1         & 0.4$\pm$0.1        &      8.22         \\
    &        &   1.9$\pm$0.7 &        36.5$\pm$0.1        &  0.5$\pm$0.1         &     3.87         \\
    &   32.9     &  5.3$\pm$0.7 &        37.9$\pm$0.1         &  1.1$\pm$0.1        &      4.61         \\
    &       &  5.8$\pm$        0.7 &        38.7$\pm$0.1        &  0.3$\pm$0.1         &     17.2         \\
    &       &   27.1$\pm$0.7 &        39.2$\pm$0.1         & 0.5$\pm$0.1         &     55.4          \\
    &       &   3.6$\pm$0.7 &        42.4$\pm$0.1         &  10.4$\pm$0.1        &     0.321          \\
18102-1800   &   21.1$^{d}$   &    3.8$\pm$0.1        &  21.6$\pm$0.1         &   0.5$\pm$0.1  &        7.05            \\   
    &      &   1.5$\pm$0.1        &  23.1$\pm$0.1        &    0.4$\pm$0.1  &        3.09            \\  
   &        &   2.9$\pm$0.1  &        24.0$\pm$0.1 &          0.6$\pm$0.1        &      4.23          \\
    &        &  1.40$\pm$0.04 &         25.020$\pm$0.006           & 0.49$\pm$0.02   &   2.69          \\        
    &       &   1.11$\pm$0.04 &         25.640$\pm$0.007          & 0.46$\pm$0.02   &   2.28          \\
    &       &   0.19$\pm$0.03 &         27.22$\pm$0.03 &          0.51$\pm$0.1 &     0.35 \\
18134-1942     & 10.5   &     2.28$\pm$0.08 &         4.877$\pm$0.004 &          0.28$\pm$0.01  &    7.56  \\
    &   &      42$\pm$3 &        6.4$\pm$0.1        &   1.1$\pm$0.1        &      35.0          \\
     &    &     37$\pm$3 &        6.6$\pm$0.1        &  0.3$\pm$0.1          &    107.          \\
    &     &    40$\pm$3 &        7.7$\pm$0.1        &  0.4$\pm$0.1         &     83.2          \\
    &     &     8$\pm$3 &        8.3$\pm$0.1        &  0.5$\pm$0.1         &     16.7          \\
    &     &     14$\pm$3 &        8.8$\pm$0.1        &  0.4$\pm$0.1          &    31.1          \\
    &     &      3$\pm$3 &         13.3$\pm$0.1        &  0.5$\pm$0.1          &    5.64          \\
    &     &   1.6$\pm$0.3 &        14.4$\pm$0.1        &  0.5$\pm$0.1  &    3.01          \\
    &      &     7.3$\pm$0.4 &        14.7$\pm$0.1        &  0.4$\pm$0.1          &    16.4          \\
   &        &     2.5$\pm$0.4 &        15.4$\pm$0.1        &  0.4$\pm$0.1          &    5.26          \\
    &      &   1.3$\pm$0.4 &        16.1$\pm$0.1        &  0.4$\pm$0.1          &    3.20          \\
18151-1208    &   32.8     &   11.75$\pm$0.09 &         27.630$\pm$0.002 &    0.323$\pm$0.003 &    34.1          \\
    &       &   31.16$\pm$0.08 &         28.000$\pm$0.003 &     0.258$\pm$0.007 &     113.          \\
    &        &   0.57$\pm$0.05 &          28.990$\pm$0.007   &  0.30$\pm$0.03 &      1.78          \\
18159-1648    &   22.1     &   0.20$\pm$0.04  &    22.7$\pm$0.1        &       1.2$\pm$0.4        & 0.154    \\
18236-1205    &   26.2      &  0.9$\pm$0.2 &        25.1$\pm$0.1 &           1.1$\pm$0.1 &             0.746         \\
     &        &   1.8$\pm$0.2 &        25.2$\pm$0.1 &  0.5$\pm$0.1        &      3.37         \\
    &        &   3.6$\pm$0.2 &        27.2$\pm$0.1 &          0.5$\pm$0.1          &    6.19         \\
    &         &  3.0$\pm$0.2 &        28.7$\pm$0.1 &          0.5$\pm$0.1        &      5.98         \\
    &       &  0.3$\pm$0.2 &        29.9$\pm$0.1 &          0.9$\pm$0.1 &             0.341         \\
18247-1147     &   55.9    &    0.34$\pm$0.02        &  116.300$\pm$0.005 &          0.23$\pm$0.01  &    1.38 \\
   &        &  0.31$\pm$0.02 &          123.00$\pm$0.02         & 0.39$\pm$0.04  &   0.756         \\
18265-1517    &  19.0\tablefootmark{b}    &    8.83$\pm$0.06    &  14.900$\pm$0.003 &         0.256$\pm$0.002 &      32.4 \\ 
     &      &    1.44$\pm$0.09    &  15.55$\pm$0.02 &         0.53$\pm$0.04 &     2.55 \\
     &       &   6.02$\pm$0.06   &   17.250$\pm$0.002 &         0.324$\pm$0.004  &    17.4  \\
   &       &   2.33$\pm$0.06    &  18.660$\pm$0.005 &         0.38$\pm$0.01  &    5.82 \\
18316-0602    &  42.2    &    0.5$\pm$0.4        & 38.5$\pm$0.1        &   0.4$\pm$0.1 &               1.41           \\ 
    &      &    4.0$\pm$0.4         & 41.0$\pm$0.1        &   0.9$\pm$0.1 &               4.04           \\ 
    &      &     39.3$\pm$0.4         & 41.8$\pm$0.1        &   0.3$\pm$0.1 &               118.           \\ 
    &       &  0.7$\pm$0.4        & 42.5$\pm$0.1        &   0.4$\pm$0.1        &       1.87           \\
18317-0845    &  62.4      &  0.22$\pm$0.03 &          59.67$\pm$0.05  &   0.7$\pm$0.1 &              0.304          \\         
    &         &  0.27$\pm$0.04 &          63.15$\pm$0.03  &   0.48$\pm$0.06 &    0.537           \\
    &         & 0.60$\pm$0.07 &          63.890$\pm$0.005  &   0.28$\pm$0.02  &    1.99           \\
    &        & 3.5$\pm$0.1 &         64.39$\pm$0.01  &   0.96$\pm$0.04 &     3.39           \\
    &        &   0.49$\pm$0.05 &          65.020$\pm$0.005 &    0.26$\pm$0.02 &     1.75          \\
    &       &   0.18$\pm$0.02 &          65.77$\pm$0.01  &   0.22$\pm$0.02  &   0.771           \\
18372-0541    &    22.8   &     0.95$\pm$0.04 &          18.25$\pm$0.01 &          0.54$\pm$0.02  &    1.65 \\   
    &        &    2.81$\pm$0.03 &         22.900$\pm$0.001        &  0.296$\pm$0.004  &    8.91        \\   
     &       &   2.8$\pm$0.1         & 24.06$\pm$0.02        &   1.00$\pm$0.06  &    2.67        \\  
    &         & 3.7$\pm$0.1 &         24.860$\pm$0.003        &  0.433$\pm$0.007 &     8.11        \\ 
18408-0353    &   102.1\tablefootmark{b}      &  0.20$\pm$0.04 &          100.00$\pm$0.06         & 0.7$\pm$0.2 &             0.293        \\
    &         &  0.11$\pm$0.03 &          105.30$\pm$0.03         &  0.3$\pm$0.1 &             0.334         \\
   &        &   0.57$\pm$0.02 &          106.400$\pm$0.003 &          0.238$\pm$0.009 &      2.24 \\
18488+0000   &    83.2   &     2.7$\pm$0.3  &        90.2$\pm$0.1 &          0.6$\pm$0.1        &      4.17          \\                
   &       &     1.7$\pm$0.3  &        90.9$\pm$0.1 &          0.4$\pm$0.1        &      3.78         \\
   &       &    0.6$\pm$0.3 &        91.4$\pm$0.1 &          0.2$\pm$0.1        &      2.69         \\
   &       &    6.3$\pm$0.3  &        91.9$\pm$0.1 &          0.4$\pm$0.1        &      13.8         \\        
   &       &    0.27$\pm$0.02 &          94.190$\pm$0.007        &  0.24$\pm$0.02 &     1.07         \\
    &      &    0.12$\pm$0.02 &          94.69$\pm$0.02        &  0.25$\pm$0.04 &    0.478         \\
18507+0121    &   57.1    &    8.62$\pm$0.02 &         55.660$\pm$0.003  &   0.394$\pm$0.003 &     20.6          \\
   &       &    0.17$\pm$0.01 &          57.860$\pm$0.003  &    0.20$\pm$0.01  &   0.766          \\
   &       &    0.93$\pm$0.08 &          59.9$\pm$0.1        &   0.4$\pm$0.1 &               2.05          \\
   &       &    1.8$\pm$0.08 &         62.5$\pm$0.1 &           0.6$\pm$0.1 &               2.73          \\
18517+0437   &   43.7    &    1.4$\pm$0.1 &        39.64$\pm$0.03 &           1.0$\pm$0.1 &              1.33          \\
   &         &    138.1$\pm$0.3  &        41.180$\pm$0.003 &          0.524$\pm$0.003 &     248.  \\
   &         &  1.56$\pm$0.2  &        42.75$\pm$0.04 &          0.8$\pm$0.1 &              1.88         \\
   &          &  3.2$\pm$0.2 &        45.34$\pm$0.05 &           1.7$\pm$0.1 &              1.81         \\
    &        &   1.67$\pm$0.08 &         45.910$\pm$0.004 &          0.23$\pm$0.01 &     6.74  \\
    &        &  0.69$\pm$0.08 &          48.22$\pm$0.02 &          0.40$\pm$0.06 &     1.62  \\
    &        &  2.35$\pm$0.08 &         51.26$\pm$0.01        &  0.50$\pm$0.02 &     4.38         \\
18532+0047    &  58.6     &   0.23$\pm$0.02 &     61.89$\pm$0.02  &   0.48$\pm$0.05 &         0.449       \\
18566+0408    &  85.1     &   0.12$\pm$0.02 &          78.41$\pm$0.02         & 0.26$\pm$0.05 &     0.447          \\
    &       &    0.31$\pm$0.02 &          79.19$\pm$0.01          & 0.33$\pm$0.03  &   0.882          \\
   &       &   1.12$\pm$0.02 &         79.680$\pm$0.004        &  0.37$\pm$0.01 &      2.80         \\
   &      &    1.56$\pm$0.07 &         83.7$\pm$0.1          & 0.3$\pm$0.1 &              4.72          \\        
   &       &    1.39$\pm$0.07 &         84.3$\pm$0.1         &  0.6$\pm$0.1 &              2.09          \\        
   &       &    0.96$\pm$0.07 &          84.9$\pm$0.1          & 0.3$\pm$0.1 &              2.65          \\        
   &       &    1.31$\pm$0.07 &         86.3$\pm$0.1         &  0.6$\pm$0.1        &      2.18          \\        
   &      &     0.23$\pm$0.07 &          87.7$\pm$0.1         &  0.7$\pm$0.1        &     0.302          \\        
19230+1341   &   39.6      &    6.69$\pm$0.01   &   35.800$\pm$0.001        &  0.658$\pm$0.003   &   9.56 \\
    &         &    16.10$\pm$0.05  &    36.900$\pm$0.001        &  0.867$\pm$0.003  &    17.4 \\
    &        &    1.00$\pm$0.03  &    38.390$\pm$0.003        &  0.32$\pm$0.01    &  2.97          \\
    &          &  3.43$\pm$0.05     & 39.180$\pm$0.005        &  0.93$\pm$0.02    &  3.47          \\
   &         &     0.95$\pm$0.03  &    40.65$\pm$0.01        &  0.74$\pm$0.03 &     1.20        \\
19282+1814   &  24.1       &  1.55$\pm$0.02    &  18.860$\pm$0.004   &  0.513$\pm$0.008        &  2.83     \\
19388+2357    & 34.6   &      3.3$\pm$0.4 &        36.4$\pm$0.1        &  0.7$\pm$0.1        &      4.18          \\   
    &       &   3.5$\pm$0.4 &        37.2$\pm$0.1        &  0.5$\pm$0.1         &     6.75        \\   
    &      &    8.8$\pm$0.4 &        38.2$\pm$0.1        &  0.4$\pm$0.1 &     23.0         \\   
    &        &   0.4$\pm$0.4 &        38.7$\pm$0.1        &  0.2$\pm$0.1           &   1.63        \\   
    &     &    1.1$\pm$0.4 &        39.6$\pm$0.1        &  0.4$\pm$0.1         &     2.90         \\   
20062+3550    &    0.6    &     0.49$\pm$0.05        &  -3.0$\pm$0.1        &   0.5$\pm$0.1        &      0.883 \\
    &         &   0.48$\pm$0.05        &  -2.6$\pm$0.1        &   0.3$\pm$0.1        &   1.57         \\
   &        &    1.47$\pm$0.05        &  5.9$\pm$0.1        &   0.4$\pm$0.1        &       3.75          \\
    &    0.6   &  0.83$\pm$0.05        &   6.6$\pm$0.1        &   0.4$\pm$0.1           &    1.92        \\
20126+4104    &  -3.9    &     1.49$\pm$0.03        & -8.393$\pm$0.005  &   0.56$\pm$0.02 &     2.52         \\
     &      & 4.6$\pm$0.9        & -7.7$\pm$0.1        &   0.4$\pm$0.1        &       11.0         \\
    &        & 4.9$\pm$0.9        & -6.5$\pm$0.1         &  0.5$\pm$0.1         &      9.09           \\
   &       &    18.5$\pm$0.9        & -6.6$\pm$0.1        &    1.3$\pm$0.1         &      13.4           \\
    &         &  12.2$\pm$0.9 &        -6.2$\pm$0.1         &  0.3$\pm$0.1          &     41.4           \\
    &      &  1.2$\pm$0.9 &        -4.9$\pm$0.1         &  0.4$\pm$0.1          &     2.68           \\
20350+4126    &    --2.5\tablefootmark{b}    &  0.20$\pm$0.03  &   -10.93$\pm$0.08        &  0.9$\pm$0.2          &   0.206          \\
   &      &  0.09$\pm$0.02   &  -4.07$\pm$0.04        &  0.31$\pm$0.07  &   0.268          \\
  21306+5540*  &  --71.1$^{\rm f}$ & 0.13$\pm$0.02  &   -78.27$\pm$0.05  &     0.6$\pm$0.1        &      0.22 \\
                            &                      &  0.38$\pm$0.02  &     -73.33$\pm$0.01  &     0.36$\pm$0.02 &      1.01 \\
                           &                       &  0.49$\pm$0.03  &     -71.04$\pm$0.02  &    0.69$\pm$0.06  &     0.66 \\
                           &                   &    0.47$\pm$0.03   &     -69.90$\pm$0.01   &     0.46$\pm$0.04  &     0.95 \\
                           &                   &    0.20$\pm$0.03  &     -68.66$\pm$0.08   &     1.0$\pm$0.2         &     0.2  \\  
22272+6358   & -9.9    &    0.46$\pm$0.02        & -12.880$\pm$0.008        &  0.30$\pm$0.02   &   1.43        \\
    &     &    0.37$\pm$0.02        & -12.40$\pm$0.02        &  0.41$\pm$0.03   &   0.853         \\  
   &     &    20.78$\pm$0.02        & -11.010$\pm$0.003        &  0.367$\pm$0.003   &   53.1          \\ 
   &     &    6.03$\pm$0.02        & -10.270$\pm$0.001        &  0.372$\pm$0.002    &  15.2         \\
22506+5944    & -51.5  &        0.17$\pm$0.04    & -55.6$\pm$0.1         &  0.9$\pm$0.3        &      0.172           \\
    &     &     0.81$\pm$0.02        &  -53.880$\pm$0.006  &    0.44$\pm$0.01  &     1.72  \\         
    22543+6145*  &  --11.1\tablefootmark{c}  &   71$\pm$1 &        -4.7$\pm$0.1        &   0.6$\pm$0.1 &               147 \\
                              &                            & 129$\pm$1 &        -4.1$\pm$0.1        &   0.3$\pm$0.1  &               361 \\          
                              &                           & 101$\pm$1  &         -3.8$\pm$0.1         &  0.2$\pm$0.1        &       394 \\
                              &                            & 156$\pm$1 &   -2.5$\pm$0.1        &   0.5$\pm$0.1        &       296 \\  
                             &                           &  30$\pm$1  &        -1.8$\pm$0.1         &  0.3$\pm$0.1        &       92 \\  
\hline
\end{longtable}
\tablefoot{
\tablefoottext{*}{sources rejected by Palla et al.~(\citeyear{palla});}
\tablefoottext{a}{from Molinari et al.~(\citeyear{mol96}) except when differently specified;}
\tablefoottext{b}{from \AMM\ observations performed with the Nobeyama 45-m telescope simultaneously to
the \METH\ maser observations;}
\tablefoottext{c}{from CS (2--1) observations, Bronfman et al.~(\citeyear{bronfman});}
\tablefoottext{d}{from Sridharan et al.~(\citeyear{sridharan});}
\tablefoottext{e}{from Richards et al.~(\citeyear{richards});}
\tablefoottext{f}{from Sunada et al.~(\citeyear{sunada}) and references therein.}
}
}

\addtocounter{table}{3}

\longtab{3}{
\begin{longtable}{cccccc}
\caption{\label{par_classII_low} Same as Tab.~\ref{par_classII_high} for \low\ sources.} \\
\hline \hline
Source & \Vlsr\ \tablefootmark{a} & $\int F_\nu {\rm d}V$  &    $V$     &      $\Delta V$    &     $F_{\rm peak}$  \\
        & (km s$^{-1}$)  &  (Jy km s$^{-1}$)  & (km s$^{-1}$)  &  (km s$^{-1}$)  &  (Jy)  \\
\hline
\endfirsthead
\caption{continued.}\\
\hline\hline
Source & \Vlsr\ \tablefootmark{a} & $\int F_\nu {\rm d}V$  &    $V$     &      $\Delta V$    &     $F_{\rm peak}$  \\
        & (km s$^{-1}$)  &  (Jy km s$^{-1}$)  & (km s$^{-1}$)  &  (km s$^{-1}$)  &  (Jy)  \\
\hline
\endhead
\hline
\endfoot
05137+3919    &  -25.4    &    0.21$\pm$0.06        & -16.2$\pm$0.1         &   0.6$\pm$0.2    &     0.3           \\ 
06099+1800* &   7.1\tablefootmark{f} &   1.8$\pm$0.9        &     1.9$\pm$0.1   &      0.6$\pm$0.1          &  2.8  \\
                         &                                 &    5.1$\pm$0.9        &     4.0$\pm$0.1    &     0.3$\pm$0.1          &  15.9 \\
                       &                                  &    28.7$\pm$0.9         &    4.6$\pm$0.1     &   0.4$\pm$0.1        &   63.6 \\
                      &                                   &  15.6$\pm$0.9         &    5.4$\pm$0.1   &     0.4$\pm$0.1         &  35.4  \\
17450-2742   &  -16.9     &   0.39$\pm$0.02    & -12.15$\pm$0.01  &   0.37$\pm$0.03        &  0.986 \\        
18024-2119    &   0.5    &    25$\pm$4    &  -5.33$\pm$0.03        &  0.42$\pm$0.08    &  56.0          \\
    &       &    67$\pm$3            &  -4.29$\pm$0.07        &  0.8$\pm$0.2          &   78.9          \\
    &       &    28$\pm$2    &  -3.81$\pm$0.02        &  0.32$\pm$0.05  &     83.9 \\         
    &       &    39$\pm$2           &   -3.22$\pm$0.03        &  0.50$\pm$0.07  &    73.4        \\  
    &       &    27$\pm$2    &  -5.3$\pm$0.1        &  0.5$\pm$0.1         &    55.4         \\
    &        &   48$\pm$2    &  -4.4$\pm$0.1        &  0.6$\pm$0.1         &    74.2        \\
    &        &   51$\pm$2    &  -3.8$\pm$0.1        &  0.5$\pm$0.1         &    105.        \\
    &        &   33$\pm$2    &  -3.2$\pm$0.1        &  0.4$\pm$0.1         &    76.5        \\
    &       &    5$\pm$2   &   -2.6$\pm$0.1        &  0.6$\pm$0.1         &    8.65        \\
    &       &    6.2$\pm$0.3    &     -0.827$\pm$0.004        &  0.86$\pm$0.06  &    6.75  \\         
    &        &    10.7$\pm$0.2       &    -9.90.09$\pm$0.006         &  0.62$\pm$0.01  &    16.1        \\
    &       &   54.3$\pm$0.3         &      1.891$\pm$0.004         &  1.066$\pm$0.009   &   47.9        \\
    &       &    24.4$\pm$0.1 &        2.69$\pm$0.01         &   1.29$\pm$0.03  &    17.7         \\
18144-1723    &  47.3    &    1.40$\pm$0.04  &    47.510$\pm$0.005  &         0.43$\pm$0.01  &    3.04 \\        
    &       &    8.95$\pm$0.04   &   49.020$\pm$0.003  &         0.464$\pm$0.003  &    18.1  \\        
    &       &   6.35$\pm$0.04   &   49.600$\pm$0.003  &         0.355$\pm$0.003  &    16.8 \\         
    &       &   2.18$\pm$0.03   &   50.330$\pm$0.004  &         0.464$\pm$0.007  &    4.42  \\        
    &       &   12.51$\pm$0.03   &   51.090$\pm$0.003  &         0.479$\pm$0.003  &    24.5  \\        
    &      &    1.50$\pm$0.02   &   51.900$\pm$0.003  &         0.344$\pm$0.006    &   4.10  \\        
18181-1534     &  --5.8\tablefootmark{b}     &   1.61$\pm$0.03   &  -3.740$\pm$0.002  &         0.296$\pm$0.006    &   5.1  \\        
     &       &   20.06$\pm$0.03   &  -2.999$\pm$0.003  &         0.399$\pm$0.003   &   47.2  \\         
18224-1228     &  48.8\tablefootmark{b}     &   2.14$\pm$0.02 &         38.700$\pm$0.002         & 0.394$\pm$0.005   &   5.1         \\
    &       &   0.08$\pm$0.01 &          39.60$\pm$0.02         &  0.19$\pm$0.04  &   0.401         \\
18278-1009    &  93.7   &     0.07$\pm$0.02 &          108.70$\pm$0.02 &     0.19$\pm$0.09  &   0.35  \\          
    &       &    0.45$\pm$0.02  &          109.800$\pm$0.007   &  0.35$\pm$0.02   &   1.19           \\
    &       &   0.46$\pm$0.02  &          110.900$\pm$0.005    & 0.30$\pm$0.01   &   1.46          \\
    &       &   1.7$\pm$0.2  &         115.9$\pm$0.1        &    1.1$\pm$0.1        &       1.48                    \\
    &       &    2.5$\pm$0.2  &         116.5$\pm$0.1        &   0.5$\pm$0.1          &     4.46                    \\
    &      &    5.4$\pm$0.2  &         117.2$\pm$0.1        &   0.5$\pm$0.1  &               9.98          \\         
    &      &    1.1$\pm$0.2  &         117.9$\pm$0.1        &   0.4$\pm$0.1           &    2.50                  \\
    &  93.7   &    0.4$\pm$0.2 &         118.4$\pm$0.1        &   0.3$\pm$0.1        &       1.13          \\         
    &      &    1.24$\pm$0.02   &         119.000$\pm$0.003  &    0.405$\pm$0.009  &    2.88  \\          
    &       &   0.86$\pm$0.04   &          119.600$\pm$0.007  &   0.39$\pm$0.02   &   2.08          \\ 
    &      &    0.22$\pm$0.04   &          120.20$\pm$0.05  &   0.5$\pm$0.1        &      0.381 \\          
18441-0134     &  57.6\tablefootmark{b}   &    0.22$\pm$0.07  &          78.0$\pm$0.1          &  0.2$\pm$0.1          &    1.02                  \\
    &       &    2.05$\pm$0.07  &         79.2$\pm$0.1     &    0.8$\pm$0.1         &     2.54                  \\
    &      &     3.53$\pm$0.07  &         80.9$\pm$0.1         &  1.1$\pm$0.1          &    3.04          \\
    &       &    0.36$\pm$0.02  &          83.080$\pm$0.005         &  0.26$\pm$0.01   &   1.31  \\        
    &         &   0.35$\pm$0.02  &          83.870$\pm$0.009         &  0.35$\pm$0.02   &   0.946        \\ 
18511+0146    &   56.8   &      0.12$\pm$0.02  &  59.74$\pm$0.03    &    0.42$\pm$0.08   &  0.27626   \\
18527+0301    &  76.0    &    3.6$\pm$0.3  &        70.3$\pm$0.1        &  0.6$\pm$0.1          &    5.78          \\
    &      &    2.0$\pm$0.3  &         71.9$\pm$0.1        &  0.6$\pm$0.1          &    3.00           \\
    &      &    10.1$\pm$0.3  &         72.8$\pm$0.1        &  0.7$\pm$0.1          &    14.4            \\
    &      &    9.1$\pm$0.3  &         74.3$\pm$0.1        &  0.5$\pm$0.1          &    18.2         \\
    &       &   0.8$\pm$0.3  &        76.1$\pm$0.1        &  0.4$\pm$0.1  &    2.02           \\
    &      &    1.06$\pm$0.08 &         81.4$\pm$0.1          &  0.6$\pm$0.1           &    1.55         \\
    &       &   1.32$\pm$0.08 &         82.1$\pm$0.1          &  0.4$\pm$0.1           &   3.29         \\
    &      &    2.64$\pm$0.08 &         83.9$\pm$0.1          &  0.7$\pm$0.1           &   3.78        \\
19012+0505    &   40.4    &    0.32$\pm$0.04 &     31.95$\pm$0.06  &    1.2$\pm$0.2        &   0.243  \\    
19092+0841   &   58.0     &   3.7$\pm$0.09 &         54.8$\pm$0.1         &  0.6$\pm$0.1           &   6.31         \\
    &       &    0.56$\pm$0.09 &          55.5$\pm$0.1         &  0.4$\pm$0.1           &   1.49         \\
    &       &    0.61$\pm$0.09  &          56.2$\pm$0.1          &  0.4$\pm$0.1           &   1.34         \\
    &        &   0.84$\pm$0.09 &          56.8$\pm$0.1         &  0.6$\pm$0.1           &   1.24          \\
    &        &   0.16$\pm$0.09 &          58.1$\pm$0.1         &  0.3$\pm$0.1          &   0.539          \\
    &      &    1.67$\pm$0.04 &         62.440$\pm$0.009        &  0.96$\pm$0.03 &     1.63 \\        
19120+1148   &  55.0     &   0.16$\pm$0.02 &     57.96$\pm$0.04   &          0.6$\pm$0.1  &   0.251 \\        
   &       &   0.21$\pm$0.02 &     59.00$\pm$0.01    &      0.39$\pm$0.05   &  0.519          \\
   &       &    0.15$\pm$0.03  &    62.11$\pm$0.09        &   1.0$\pm$0.2          &   0.141         \\
19186+1440   &     --    &    0.47$\pm$0.02 &  -26.670$\pm$0.007  &         0.33$\pm$0.02  &    1.33 \\        
   &         &   1.37$\pm$0.03   &  -25.800$\pm$0.008  &         0.79$\pm$0.02   &   1.62 \\        
   &         &   0.18$\pm$0.05  & -23.99$\pm$0.08  &         0.9$\pm$0.5         &   0.185          \\
   &         &   0.92$\pm$0.07  & -14.86$\pm$0.03  &         0.78$\pm$0.06  &     1.11 \\            
    &        &   0.75$\pm$0.06 &  -14.370$\pm$0.006   &         0.37$\pm$0.01   &   1.90 \\        
    &        &   2.52$\pm$0.03   &  -13.380$\pm$0.003  &         0.677$\pm$0.008   &   3.50  \\            
    &        &   3.15$\pm$0.03   &  -12.110$\pm$0.003  &         0.698$\pm$0.006    &   4.24  \\             
    &        &   2.97$\pm$0.03   &  -10.750$\pm$0.004  &         1.00$\pm$0.01   &   2.80  \\           
22187+5559    &   --45.2\tablefootmark{e}     &   0.09$\pm$0.03        & -11.44$\pm$0.08   &  0.6$\pm$0.3  &         0.133        \\
\hline
\end{longtable}    
}

\begin{table*}
\begin{center}
\caption[]{Parameters of the 44 GHz \METH\ maser (Class I) detected with the Nobeyama 45-m
telescope for both \high\ (the IRAS name begins with H) and \low\ (the IRAS name
is begins with L) sources.}
\label{par_classI}
\begin{tabular}{cccccc}
\hline \hline
Source\tablefootmark{a} & \Vlsr\ & $\int F_{\nu} {\rm d}V$  &    $V$     &      $\Delta V$    &     $F_{\nu}^{\rm peak}$  \\
        & (km s$^{-1}$)  &  (Jy km s$^{-1}$)  & (km s$^{-1}$)  &  (km s$^{-1}$)  &  (Jy)  \\
\hline
 H05168+3634           &  --15.1  &   1.12$\pm$0.2   &      --4.9$\pm$0.2    &      1.3$\pm$0.4     &    0.80           \\
 H05480+2545          &   --9.3\tablefootmark{b}  & 3.3$\pm$0.1 &     --9.79$\pm$0.02 &    0.77$\pm$0.05 &     4.08 \\    
 H06103+3030          &  15.6   &  2.7$\pm$0.6       &   16.4$\pm$0.7      &        5$\pm$1.5       &  0.51    \\
 H18089-1732          &   32.9 &  10.9$\pm$0.4     &     31.94$\pm$0.01 &    0.86$\pm$0.04 &     11.8  \\     
 H18151-1208          &   32.8 &  2.2$\pm$0.2   &       30.5$\pm$0.1      &   1.5$\pm$0.3      &   1.34    \\
           &    &  2.0$\pm$0.4    &      33.8$\pm$0.2       &   2.1$\pm$0.5      &   0.89    \\
 H18159-1648          &   22.1 &  4.0$\pm$0.6      &    13.2$\pm$0.1      &   0.9$\pm$0.1       &   3.95    \\
           &    &  10.9$\pm$0.6      &    19.5$\pm$0.1      &    1.0$\pm$0.1        &   10.01    \\
           &    &  15.8$\pm$0.6     &    20.9$\pm$0.1     &   1.0$\pm$0.1      &    6.97    \\
           &    &  31.4$\pm$0.6     &    22.9$\pm$0.1      &   0.9$\pm$0.1      &    32.11    \\
           &    &  18.5$\pm$0.6    &    23.9$\pm$0.1     &   0.6$\pm$0.1      &    27.9    \\
 H18316-0602          &   42.2 &  6.6$\pm$0.1  &    42.060$\pm$0.008  &    0.62$\pm$0.02  &     9.88  \\     
           &    &  1.6$\pm$0.2     &     43.10$\pm$0.04  &   0.70$\pm$0.09   &  2.07    \\
           &    &  2.9$\pm$0.4      &    44.52$\pm$0.09    &   1.6$\pm$0.3      &   1.67     \\
 H18360-0537          &   102.3  &  6.7$\pm$0.4      &    102.4$\pm$0.1    &   2.83$\pm$0.2        &  2.23    \\
           &     &  1.6$\pm$0.4    &      106.4$\pm$0.3     &     1.9$\pm$0.7      &   0.74  \\ 
 H18507+0121          &   57.1 &  10.0$\pm$0.4    &     60.33$\pm$0.04 &      1.3$\pm$0.1 &         7.00 \\    
 H18517+0437          &   43.7 &  3.1$\pm$0.2     &     43.97$\pm$0.05 &      1.1$\pm$0.1 &         2.70   \\  
 H19043+0726          &   58.9  &  2.5$\pm$0.2     &    58.69$\pm$0.03   &  0.65$\pm$0.09 &     3.52    \\ 
           &     &  1.1$\pm$0.2    &      59.89$\pm$0.08 &    0.7$\pm$0.2     &    1.31    \\
 H19088+0902          &   59.6  &  10$\pm$1      &    58.67$\pm$0.07   &   2.2$\pm$0.2      &   2.0     \\
 H19388+2357          &   34.6 &  1.8$\pm$0.4   &       35.4$\pm$0.2     &    1.6$\pm$0.4     &    1.07    \\
 H20050+2720          &   6.4   &  2.0$\pm$0.2      &    6.6$\pm$0.1      &   1.8$\pm$0.3      &   1.09     \\
 H20062+3550          &   0.6 &  2.5$\pm$0.2     &    0.60$\pm$0.06 &     1.0$\pm$0.1  &        2.36  \\   
 H20126+4104          &   -3.9 &  8.9$\pm$0.4    &     -2.43$\pm$0.04 &     2.0$\pm$0.1  &        4.17    \\ 
 H20188+3928          &   1.5   &  2$\pm$1   &      3$\pm$1    &    3$\pm$2         &   0.76  \\
 H21391+5802          &    0.4  &  4.2$\pm$0.2    &    -0.49$\pm$0.02 &    0.73$\pm$0.08 &     5.33  \\   
           &      &  5.4$\pm$0.4    &      6.6$\pm$0.1   &   2.3$\pm$0.3      &    2.25     \\
 H22506+5944          &   -51.5 &  3$\pm$1   &     -51.3$\pm$0.1  &        1.6$\pm$0.6    &      1.81 \\
          &     &  4$\pm$2   &     -48$\pm$1  &        5$\pm$2    &      0.8 \\
 L00420+5530          &    -51.20  &   1.6$\pm$0.4  &     -48.8$\pm$0.3    &      2.0$\pm$0.5    &     0.8 \\
 L18018-2426          &    10.5  &  241.1$\pm$0.4     &     11.280$\pm$0.003 &    0.669$\pm$0.002  &    339 \\    
 L18024-2119          &   0.5   &   10.0$\pm$0.4      &   0.31$\pm$0.02 &     1.28$\pm$0.05   &   7.34    \\
           &      &  3.1$\pm$0.4       &   2.7$\pm$0.1      &   2.3$\pm$0.3      &   1.4     \\
 L18144-1723          &   47.3 &   42.1$\pm$0.2     &    48.590$\pm$0.005 &      1.54$\pm$0.01  &    25.6 \\     
 L18162-1612          &   61.8 &   5.1$\pm$0.2  &    62.80$\pm$0.03 &    0.84$\pm$0.07  &    3.2  \\     
 L18396-0431          &   97.3 &   1.0$\pm$0.2      &    97.6$\pm$0.1      &   1.0$\pm$0.3   &      1.0    \\
 L19092+0841          &   58.0 &   4.0$\pm$0.4        &  59.07$\pm$0.4       &  5.7$\pm$0.7       &  0.65     \\
 L23385+6053          &   -50.0  &  2.9$\pm$0.4      &  -50.6$\pm$0.3      &    3.2$\pm$0.7     &    0.9    \\
 \hline
\end{tabular}
\tablefoot{
\tablefoottext{a}{ ``L'' and ``H'' indicate if the source belongs to the \low\ or \high\ sample;}
\tablefoottext{b}{from \AMM\ observations performed with the Nobeyama 45-m telescope.}
}
\end{center}
\end{table*}         

\begin{table*}
\begin{center}
\caption[]{Same as Table~\ref{par_classI} for the 95 GHz \METH\ maser (Class I) detected with the Nobeyama 45-m
telescope.}
\label{tab_par95GHz}
\begin{tabular}{cccccc}
\hline \hline
Source\tablefootmark{a} & \Vlsr\ & $\int F_{\nu} {\rm d}V$ &    $V$     &      $\Delta V$    &     $F_{\nu}^{\rm peak}$  \\
        & (km s$^{-1}$)  &  (Jy km s$^{-1}$)  & (km s$^{-1}$)  &  (km s$^{-1}$)  &  (Jy)  \\
\hline
 H05480+2545  &    --9.3\tablefootmark{b}  & 3.1$\pm$0.6      &   -10.6$\pm$0.1     &    1.3$\pm$0.3      &   2.2  \\
 H18089-1732    &   32.9 &  2.7$\pm$0.4    &     31.38$\pm$0.02  &        0.45$\pm$0.05  &       5.5 \\  
    &    &  14$\pm$2      &   32.4$\pm$0.2  &       3.2$\pm$0.4   &      4.2   \\
 H18159-1648   &  22.1 &  4.7$\pm$0.8      &    20.1$\pm$0.1       &   1.6$\pm$0.4       &   2.7   \\
    &   &  7.6$\pm$0.6       &   22.56$\pm$0.06 &     1.4$\pm$0.1     &    5.2  \\   
 H18316-0602   &  42.2 &  1.1$\pm$0.2 &     41.63$\pm$0.05 &    0.5$\pm$0.1 &        2.1 \\  
 H18360-0537   &  102.3 &  13.2$\pm$0.8      &   102.5$\pm$0.1      &  4.1$\pm$0.4       &   4.0    \\ 
 H20126+4104  & -3.9  &  6.0$\pm$0.4       &  -2.75$\pm$0.08    &  2.3$\pm$0.2       &   2.5   \\
 H21391+5802  &   0.4   & 2.5$\pm$0.6      &   12.1$\pm$0.1        &  1.2$\pm$0.5      &   2.0    \\
 H22506+5944  &  -51.5 &  5$\pm$1    &     -47.1$\pm$0.2      &    2.4$\pm$0.6    &     2.0   \\
 L18018-2426    &  10.5  & 33.5$\pm$0.4    &     10.800$\pm$0.004 &    0.622$\pm$0.009 &    50  \\  
 L18144-1723   &   47.3 &  26.4$\pm$0.6     &     47.91$\pm$0.03    &  2.07$\pm$0.08  &    12   \\
 L18396-0431  &  97.3  & 3.8$\pm$0.4    &     50.37$\pm$0.08  &    1.24$\pm$0.2   &       2.9   \\
    &    & 14$\pm$2       &   57.4$\pm$0.6    &      7$\pm$1     &    1.9   \\
     &   & 4$\pm$1      &    61.5$\pm$0.1   &       1.3$\pm$0.4  &        2.5   \\
    &   &  8.0$\pm$0.8   &     71.3$\pm$0.2   &        3.4$\pm$0.6 &         2.2   \\
     &    &  1.6$\pm$0.2     &     130.50$\pm$0.08 &    0.7$\pm$0.1 &        2.1 \\ 
\hline
\end{tabular}
\tablefoot{
\tablefoottext{a}{``L'' and ``H'' indicate if the source belongs to the \low\ or \high\ sample;}
\tablefoottext{b}{from \AMM\ observations performed with the Nobeyama 45-m telescope.}
}
\end{center}
\end{table*}         

\end{document}